 \documentclass[prd,aps,showpacs,preprintnumbers,amsmath,amssymb,nofootinbib,preprintnumbers]{revtex4}

\usepackage{graphicx}
\usepackage{epsf}
\usepackage{amsmath}
\usepackage{epstopdf}
\usepackage{bm}
\usepackage{color}
\usepackage{tabularx}
\usepackage{enumitem}
\usepackage{float}
\usepackage{array,booktabs}
\usepackage{footnote}
\usepackage{threeparttable}
\usepackage{graphicx}
\usepackage{hyperref}
\usepackage{amssymb,epsf}
\usepackage{latexsym}
\usepackage{epstopdf}
\usepackage{epsfig}
\usepackage{subfigure}

\usepackage[normalem]{ulem}
\useunder{\uline}{\ul}{}

\begin{document}

 \title{Thermodynamic stability of a new three dimensional regular black hole}
 \author{
 S. H. Hendi$^{1,2,3}$\footnote{email address: hendi@shirazu.ac.ir},
 S. Hajkhalili$^{1,2}$\footnote{email address: S.hajkhalili.@gmail.com} and
 S. Mahmoudi$^{1,2}$\footnote{email address: S.mahmoudi@shirazu.ac.ir} }

 \affiliation{
 $^1$Department of Physics, School of Science, Shiraz University, Shiraz 71454, Iran \\
 $^2$Biruni Observatory, School of Science, Shiraz University, Shiraz 71454, Iran \\
 $^3$Canadian Quantum Research Center 204-3002 32 Ave Vernon, BC V1T 2L7 Canada }

\begin{abstract}
A new model of the regular black hole in $(2+1)-$dimensions is
introduced by considering an appropriate matter field as the
energy-momentum tensor. First, we propose a novel model of
$d$-dimensional energy density that in $(2+1)-$dimensions leads to
the existence of an upper bound on the radius of the event horizon
and a lower bound on the mass of the black hole which are
motivated by the features of astrophysical black holes. According
to these bounds, we introduce an admissible domain for the event
horizon radius, depending on the metric parameters. After
investigation of geometric properties of the obtained solutions,
we study the thermal stability of the solution in the canonical
ensemble and find that the regular black hole is thermally stable
in the mentioned admissible domain. Besides, the Gibbs free energy
is calculated to examine the global stability of the solution.
\end{abstract}

\maketitle

\section{Introduction}

Nowadays, the theory of General Relativity (GR) still well passes
most gravitational tests. Without a doubt, the black hole is one
of the most interesting predictions of Einstein's theory of GR. In
recent years, these objects have been moved into the focus of
observational and theoretical studies in gravity and cosmology due
to the detection of
gravitational waves emitted from the merger of two black holes \cite%
{LIGOScientific:2016aoc, LIGOScientific:2016sjg, LIGOScientific:2017bnn} and
the first picture of an event horizon by the event horizon telescope
collaboration \cite{EventHorizonTelescope:2019dse}.

Roughly speaking, a special characteristic of the first known
exact black hole solutions in GR is the existence of a region of
spacetime in which the laws of physics break down \cite{penrose}.
This region, where just possible to enter while impossible to
escape from in a classical point of view is delimited by the
so-called event horizon (see a review \cite{Joshi:2013xoa} and
references therein). Nevertheless, regarding the theory of GR and
its black hole solutions, there are two theoretical difficulties:
the singularity problem and the problem of its quantization.
According to the singularity theorems, Penrose and Hawking showed
that, mathematically, singularities are unavoidable in GR
\cite{PH1,PH2,PH3,PH4,PH5}. However, regarding the Hawking
radiation of black holes, the existence of singularities is more
complicated due to their relations with the information paradox.
As a result, it is extensively believed that since these
singularities are created by classical theories of gravity from
the pure mathematical viewpoint, they are not physical objects and
they cannot exist in nature, and therefore, non-singular solutions
and their properties are well-motivated topics accordingly. For
instance, new approaches to avoid singularities were suggested in
some modified gravity theories \cite{NSBHs1,NSBHs2,NSBHs3,NSBHs4}
and nonlinear electrodynamics \cite{NLEDNSBHs}.

The main motivation for the appearance of regular solutions comes
from the fact that most physicists and philosophers believe that
singularities are not reasonable possibilities for physical
manifestation in our real world. This opinion is supported by an
indication that classical general relativity cannot be valid at
all scales \cite{PlanckStars1,PlanckStars2}, and therefore, at
high energy scales, the quantum effects should be considered. As a
consequence of loop quantum gravity, for instance, the pressure of
quantum fluctuations may counterbalance the gravitational collapse
of a typical supermassive star to avoid singularity formation. So,
a dense central core can be formed inside a black hole whose
functional density modeling would be
interesting. In this way, the Sakharov's \cite{sakharov} and Gliner's \cite%
{Gliner} quantum arguments suggest that matter source with a de
Sitter (dS) core at the center of the spacetime does not include
the spacetime singularities. According to this idea, the first
model of a regular (i.e non-singular) black hole was proposed by
Bardeen\cite{Bardeen}. In this model, the solution has a dS core
and the singularity is avoided by considering a collapse of a
charged matter with a charged matter core inside the black hole
instead of its singularity. After that, a large class of regular
black hole solutions was found based on the idea that singularity
could be replaced by a regular distribution of matter
\cite{Ayon-Beato:1998hmi, Bronnikov:2000vy, Dymnikova:2004zc,
Novello:1999pg, Culetu:2014lca, Balart:2014cga, Hayward:2005gi,
Azreg-Ainou:2014pra,Sajadi}. It is worth mentioning that although
the strong energy condition (SEC) may be violated by all regular
black hole solutions with spherical symmetry, the dominant energy
condition (DEC), as well as weak energy condition (WEC), are
satisfied in most cases \cite{Zaslavskii:2010qz}.

Regular black holes can be constructed in various circumstances. A
particular kind of non-singular black holes in which the
singularity is replaced by a dS core with regular geometry around
the origin has been proposed by Nicolini et al.
\cite{Nicolini:2005vd}. Dymnikova has found a different type of
regular black holes with a dS core inside the
horizon, smoothly connecting to a Schwarzschild as the outer geometry \cite%
{Dymnikova:1992ux}. Another interesting attempt is to construct
regular black hole solutions including matter fields in the
energy-momentum tensor (see for e.g, \cite{Hayward:2005gi,
Dymnikova:1992ux, Spallucci:2017aod,Aros:2019auf, Aros:2019quj,
Babichev:2020qpr}). Furthermore, models including nonlinear
electrodynamics or scalar fields
have been successful in the construction of non-singular solutions \cite%
{Ayon-Beato:1999kuh,Ayon-Beato:1999qin, Burinskii:2002pz,
Ma:2015gpa}.

It is also possible to have regular black holes in
lower-dimensional spacetime. Generally, models of gravity in $(2 +
1)-$dimensions have become an active field of research in recent
years. Due to the lack of the Newtonian limit and propagating
degrees of freedom \cite{Carlip:1995qv}, the general idea was that
Einstein gravity is a topological theory in $(2+1)-$ dimensional
spacetime \cite{Welling:1995er} and there is no black holes in
this geometry. However, the discovery of the famous BTZ black hole
\cite{Banados:1992wn, Banados:1992gq} aroused great interest in
the study on these objects \cite{Chan:1994qa, Zaslavsky:1994dx,
Martinez:1999qi,Yamazaki:2001ue, Gurtug:2010dr, Hendi:2015bba,
Tang:2019jkn}. The main motivation to study the Einstein gravity
in $(2+1)-$ dimensional spacetime is that it provides a simpler
framework to understand many conceptual features of $(3+1)-$
dimensional Einstein gravity \cite{QGravity}. In fact, the
simplicity of the equations of motion in three dimensional
geometry has led to being studied these models as a toy model to a
better understanding of some problems of its $(3+1)-$dimensional
counterpart and assist us to comprehend some conceptual issues
regarding the quantum gravity and string theory
\cite{Carlip:1995qv, Witten:1998zw}. Moreover, since the black
hole properties at the quantum level in $(3+1)-$ dimensional
gravity has remained as a mystery, the $(2+1)-$dimensional black
holes could provide a good relatively simple laboratory to examine
the features of black holes in higher dimensions and to find the
deeper insight into the general facets of black hole physics.
Another major motivation is related to the AdS/CFT correspondence.
The study of the near horizon of the $3-$dimensional black holes
has helped us to explore some conceptual aspects of the AdS/CFT
duality \cite{Witten:2007kt}. {In \cite{Birmingham:2001pj} the
authors have shown the coincidence between the quasinormal modes
in this geometry and poles of the correlation function in the dual
CFT which gives  evidence for $\text{AdS}_3/\text{CFT}_2$. Also,
one dimensional holographic superconductors can be explored in the
background of $(2+1)-$ dimensional black hole \cite{Ren:2010ha,
Liu:2011fy, KordZangeneh:2017zyy}.}
 An additional reason that motivates
us to study the three dimensional gravity is that the
investigation of the $(2+1)-$dimensional black holes' physical
properties has enhanced our comprehension of the gravitational
systems and their interactions in lower dimensions
\cite{Witten:2007kt}. { Another motivate for examining this
geometry is that all the characteristics of a configuration in a
given $d-$dimensions are not necessarily transferred to its lower
ones and a system in lower dimensions can exhibit different
features. For instance, it is shown that the charge term of the
lapse function in the Ricci flat $d-$dimensional
Reissner-Nordstr\"{o}m solutions is proportional to $r^{-(2d-6)}$
that is a constant term in $(2+1)-$dimensions. It indicates that
higher-dimensional Reissner-Nordstr\"{o}m black holes reduce to
uncharged solutions in three dimensions. While we know that there
is a logarithmic charge term in the lapse function of
Einstein-Maxwell system in three dimensions
\cite{Hendi:2010px}.}\par
 The study of BTZ black holes in the
noncommutative spacetime provides the possibility of the existence
of gravitational Aharonov Bohm effect is a case in point
\cite{Anacleto:2014cga}. Also, the possibility of mimicking the
BTZ black hole properties in higher dimensions has been studied in
\cite{Hendi:2010px, Ghosh:2011tt}. Besides, the authors in
\cite{Yamazaki:2001ue, Hendi:2015uia} have investigated the
existence of the $(2+1)-$ dimensional solutions in the presence of
the nonlinear electrodynamics. Three dimensional black holes not
only exist in the context of Einstein's gravity, but also the
modified gravities such as Lifshitz gravity
\cite{Ayon-Beato:2009rgu}, dilatonic gravity \cite{Chan:1994qa,
Hendi:2017mgb},  massive gravity \cite{Hendi:2016pvx},  gravity's
rainbow \cite{Hendi:2016wwj}, massive gravity's rainbow
\cite{Hendi:2016hbe} have similar solutions.\par

In this paper, we propose a new model of the regular black hole in
{ $d-$dimensions} and {then, as a special case, we} investigate
the geometric and thermodynamic properties of the
{$(2+1)-$dimensional} solution. To this end, the conventional
approach is to start from the energy-momentum tensor of a given
matter field and then to solve the field equation. On the other
hand, it is feasible to take the opposite way that the regular
black hole solution can be first constructed based on the
requirement of divergence-free curvature and then, the
corresponding energy-momentum or the corresponding matter field
will be driven
\cite{Hayward:2005gi,Balart:2014cga,Balart:2014jia,Fan:2016hvf,
Estrada:2020tbz}. Here, we will proceed according to latter
approach.

The paper is organized as follows. In the next section, we
introduce our proposed new model of the regular black hole and
study the physical properties of the solution. Section
\ref{thermo} is devoted to the investigation of thermodynamics
properties of the solution. We calculate thermodynamic quantities
and examine the first law of thermodynamics. We also explore
local/global thermodynamic stability in the canonical grand
canonical ensemble. {In the concluding remarks part, we give a
summary and conclusion. Finally, in Appendix \ref{appendix1}, we
provide a brief introduction on the curvature singularity-free
models of black holes. }


\section{New model of regular black hole and its exact solution\label{new
model}}

As we mentioned before, there are various motivations to find
appropriate solutions of Einstein field equations that describe
regular black hole. In this section, we suggest a new model of
energy density in {$d-$dimensions}, based on the inclusion of
suitable matter field in the energy-momentum tensor, and show that
it leads to an interesting {$(2+1)-$dimensional} regular black
hole solution.

Although one can introduce different ad hoc models of energy
density to have a regular black hole, we should consider
reasonable criteria to veto unphysical models. In order to have a
well-defined physical solution, the suggested energy density must
have a certain behavior at the origin as well as at infinity.
Strictly speaking, the energy density must be positive,
continuously differentiable and have a finite single maximum at
the origin to avoid singularity. Besides, to guarantee a
well-defined asymptotic behavior, the energy density should be a
decreasing function of radial coordinate to vanish at spatial
infinity, i.e. $\rho=0$ as $r \to \infty$ \cite{Aros:2019quj}
{(see the appendix \ref{appendix1} for more details)}.

Keeping in mind the mentioned criteria, we can suggest the
following energy density in $d-$dimensions
    \begin{equation}\label{rho1}
    \rho(r) =\, \frac{d-1}{\Omega_{d-2}}\,\frac{k-1}{2\,L^{d-2}} (1+\frac{r^{d-1}}{2\,L^{d-2}M})^{-k},
    \end{equation}
where $k \geq 2$ is an integer number, the constant parameter $L$
is a regulator with square dimension of length which should be
positive to make sure the positivity of the energy density at the
origin and $M$ is the dimensionless mass parameter.  We should
mention that the demand for the absence of singularity in the
energy density necessitates considering a positive definite mass
parameter, $M$. As is discussed in appendix \ref{appendix1}, our
proposed model of energy density can be considered as a
generalization of the model that Estrada and Aros proposed for
regular black holes.

Here, we are interested in studying the $(2+1)-$ dimensional
solution and properties of the higher dimensional ones will be
explored in future works. Based on Eq. \eqref{rho1}, energy
density in $(2+1)-$ dimensions will reduce to
\begin{equation}  \label{rho}
\rho(r) = \frac{k-1}{2\pi\,L}\,\left(1+\frac{r^2}{2LM}\right)^{-k},
\end{equation}
To derive an exact black hole solution corresponding to {the
above} energy density, we conventionally start from the Einstein
field equations which in $(2+1)-$dimensions is given by
\begin{equation}
G_{\,\nu }^{\mu }+\Lambda \delta _{\,\nu }^{\mu }=8\pi \mathbb{G}T_{\,\nu
}^{\mu },  \label{Einstein}
\end{equation}%
where $G_{\,\nu }^{\mu }$ and $T_{\,\nu }^{\mu }$ are,
respectively, the Einstein tensor and a second rank symmetric
tensor of the energy-momentum. Also, $\Lambda =-l^{-2}$ is the
cosmological constant which is related to the anti de Sitter (AdS)
radius, $l$, and $\mathbb{G}$ is dimensionless gravitational
constant (we consider $c=1$ throughout this paper). We assume a
spherically symmetric spacetime describing by the following metric
ansatz
\begin{equation}
g_{\mu \nu }=\mbox{diag}\big[-f(r),f(r)^{-1},r^2\big]. \label{line
element}
\end{equation}

Here, we consider a three dimensional energy-momentum tensor
describing a non-trivial anisotropic fluid with different radial
and tangential pressures
\begin{equation}\label{emtensor}
T_{\,\nu }^{\mu }=\mbox{diag}\big[-\rho (r),P_{r}(r),P_{\phi
}(r)\big],
\end{equation}%
where it is easy to show that one has to consider $P_{r}(r)=-\rho
(r)$ due to the consistency with the Einstein field equations.
In addition, taking into account Eq.
(\ref{rho}) and solving the $\phi \phi $ component of the Einstein
field equation, one can obtain the functional form of tangential
pressure
\begin{equation}
P_{\phi }=\big(\frac{2kr^2}{2LM+r^2}-1\big)\,\rho.
\label{Pt}
\end{equation}

We should note that although we suggest the energy density model,
the functional form of the pressures do not depend on the metric
function. In other words, we read the field equations
(\ref{Einstein}) from right to left. Before obtaining the metric
function, it is worth discussing the asymptotic behavior of energy
density and pressures. It is easy to find that
\begin{eqnarray*}
\lim_{r\rightarrow 0}\rho (r) &=&-\lim_{r\rightarrow
0}P_{r}(r)=-\lim_{r\rightarrow 0}P_{\phi }=\frac{k-1}{2\pi \,L}, \\
\lim_{r\rightarrow \infty }\rho (r) &=&-\lim_{r\rightarrow \infty
}P_{r}(r)=-\lim_{r\rightarrow \infty }P_{\phi }=0,
\end{eqnarray*}%
where confirm that the nonzero components of the suggested
energy-momentum tensor are finite values near the origin and they
vanish at spatial infinity. So, the matter distribution does not
affect the asymptotic behavior of the solution which is
{characterized} by the cosmological constant. Besides, unlike the
radial pressure (energy density) which is a smooth function of
radial coordinate, tangential pressure enjoys a root at finite
radius (${r=}\sqrt{\frac{2LM}{2k-1}}$) and an extremum point at ${r=}\sqrt{%
\frac{6LM}{2k-1}}$.

Regarding the functional form of nonzero components of the
energy-momentum tensor, we are in a position to obtain the metric
function. For the line element \eqref{line element}, we arrive at
the following equation by combining the $tt$ and $rr$ components
of the Einstein field equations
\begin{equation}
\frac{1}{2r}\,\frac{df}{dr}\,+\,\Lambda \,=\,-8\,\pi \,\mathbb{G}\,\rho ,
\label{equation}
\end{equation}%
with the following traditional ansatz%
\begin{equation}
f(r)=1-8\mathbb{G}m(r)-\Lambda r^{2},  \label{ansatz}
\end{equation}%
where the mass $m(r)$ is related to the energy density as%
\begin{equation}
\frac{d}{dr}m(r)=2\pi r\rho .
\end{equation}%
After a simple manipulation, the explicit form of the mass
function for the proposed energy density of $(2+1)-$dimensional
black hole can be rewritten as
\begin{equation}
m(r)=2\pi \int_{0}^{r}x\rho (x)dx=M\,\left[ 1-\left( 1+\,\frac{{r}^{2}}{2LM}%
\right) ^{1-k}\right] .  \label{mass function}
\end{equation}%
As it is expected, the mass function tends to its finite maximum
value at infinity ($m(r)\longrightarrow M$ as $r\rightarrow \infty
$) which confirms its well-defined asymptotic behavior. Now, by
substituting \eqref{mass function} into \eqref{ansatz}, the
solution will take the following analytical form
\begin{equation}
f(r)=\,1-8\,\mathbb{G}M\left[ 1-\left( 1+\frac{{r}^{2}}{2LM}\right) ^{1-k}%
\right] +{\frac{{r}^{2}}{{l}^{2}}}, \label{f metric}
\end{equation}%
which its asymptotic behavior is AdS
\begin{equation*}
\left. f(r)\right\vert _{r\rightarrow \infty }=1-8\,\mathbb{G}\,M+\frac{r^{2}%
}{l^{2}}+\mathcal{O}\left( r^{2-2k}\right) .
\end{equation*}%
In addition, one can find the regularity of the metric function near the
origin as
\begin{equation}
\left. f(r)\right\vert _{r\rightarrow 0}=1+\left[ -{\frac{4\mathbb{G}\left(
k-1\right) }{L}}+\frac{1}{l^{2}}\right] {r}^{2}+{\frac{\mathbb{G}k\left(
k-1\right) }{{L}^{2}M}}{r}^{4}+\mathcal{O}\left( {r}^{6}\right) ,  \label{f0}
\end{equation}%
which shows that our solution has no singularity at the origin. To
study the situation of the core of our regular black hole
solution, we should focus on the second term of Eq. (\ref{f0}).
Obviously, by choosing $L=4\mathbb{G}l^{2}(k-1)$, the black hole
enjoys a flat core while for $L<4\mathbb{G}l^{2}(k-1)$ $\left[
L>4\mathbb{G}l^{2}(k-1)\right] $ its core is dS [AdS].


Here, we should examine the regularity of the solutions. To do so,
one can, generally, consider some curvature invariants such as the
Kretschmann scalar, Ricci square, Ricci scalar and Weyl square. It
is worth mentioning that the Riemann tensor has six independent
components corresponding to the Ricci tensor for a general three
dimensional spacetime while for diagonal ansatz (\ref{line
element}), the nonzero components reduces to three. As a result,
the Weyl square vanishes and we find that
\begin{align*}
& \text{\textbf{Kretschmann scalar:}} \\
\mathcal{K} =&\mathcal{R}_{abcd}\mathcal{R}^{abcd}=f^{\prime
\prime
2}(r)+2\left( \frac{f^{\prime }}{r}\right) ^{2} \\
& =\frac{12}{l^{4}}-\,96\,(k-1){\frac{\mathbb{G}}{L{l}^{2}}\left[ 1-\,\frac{{%
r}^{2}}{3LM}\,\left( k-\frac{3}{2}\right) \right] \left( 1+\,{\frac{{r}^{2}}{%
2LM}}\right) ^{-k-1}}+192\,\left( k-1\right) ^{2}\frac{\mathbb{G}^{2}}{{L}%
^{2}} \\
& \qquad \qquad \times \left[ 1-\,\frac{r^{2}}{3LM}\Bigg\{2\left( k-\frac{3}{%
2}\right) +\,{\frac{{r}^{2}}{3\,LM}}\left( {k}^{2}-k+\frac{3}{4}\right) %
\Bigg\}\right] \left( 1+\,{\frac{{r}^{2}}{2LM}}\right) ^{-2\,k-2},
\end{align*}
\begin{align*}
& \text{\textbf{Ricci square:}} \\
 \qquad \mathcal{R}^{2}=&\mathcal{R}_{ab}\mathcal{R}^{ab}=\mathcal{K}-\frac{1%
}{2}\left( \mathcal{R}+3\frac{f^{\prime }}{r}\right) ^{2} \\
& =\frac{12}{l^{4}}-\,96\,(k-1){\frac{\mathbb{G}}{L{l}^{2}}\left[ 1-\,\frac{{%
r}^{2}}{3LM}\,\left( k-\frac{3}{2}\right) \right] \left( 1+\,{\frac{{r}^{2}}{%
2LM}}\right) ^{-k-1}}+192\,\left( k-1\right) ^{2}\frac{\mathbb{G}^{2}}{{L}%
^{2}} \\
& \qquad \qquad \times \left[ 1-\,\frac{r^{2}}{3LM}\Bigg\{2\left( k-\frac{3}{%
2}\right) +\,{\frac{{r}^{2}}{6\,LM}}\left( {k}^{2}-2k+\frac{3}{2}\right) %
\Bigg\}\right] \left( 1+\,{\frac{{r}^{2}}{2LM}}\right) ^{-2\,k-2},
\end{align*}
\begin{align*}
& \text{\textbf{Ricci scalar:}} \\
{\mathcal{R}}& =-f^{\prime \prime }(r)-2\left( \frac{f^{\prime }}{r}\right)  \\
&=\frac{-6}{l^{2}}+\,24\,(k-1)\frac{\,\mathbb{G}}{L}\left[
1-\,{\frac{{r}^{2}}{3LM}}\,\left( k-\frac{3}{2}\right) \right] \left( \,1+{%
\frac{{r}^{2}}{2LM}}\right) ^{-k-1}.
\end{align*}

To study more precisely, we consider their behavior in the
presence of large and small radii. No matter of the value of $k$,
all curvature scalars have finite value at infinity
\begin{equation*}
\begin{array}{cl}
\left. {\mathcal{K}}\right\vert _{{r\rightarrow \infty }}= & \frac{12}{l^{4}}+\frac{%
\alpha (k)\,\mathbb{G}\,M}{l^{2}r^{2}}\left( \frac{L\,M}{r^{2}}\right)
^{k-1}+\mathcal{O}\left( \frac{1}{r}\right) ^{{2(k+1)}}, \\
\left. \mathcal{R}^{2}\right\vert _{{r\rightarrow \infty }}= & \frac{12}{%
l^{4}}+\frac{\alpha (k)\,\mathbb{G}\,M}{l^{2}r^{2}}\left( \frac{L\,M}{r^{2}}%
\right) ^{k-1}+\mathcal{O}\left( \frac{1}{r}\right) ^{{2(k+1)}}, \\
\left. \mathcal{R}\right\vert _{{r\rightarrow \infty }}= & -\frac{6}{l^{2}}+%
\frac{\beta (k)\,\mathbb{G}\,M}{r^{2}}\left(
\frac{L\,M}{r^{2}}\right) ^{k-1}+\mathcal{O}\left(
\frac{1}{r}\right) ^{{2(k+1)}},
\end{array}%
\end{equation*}%
where $\alpha (k)$ and $\beta (k)$ are numbers which vary by changing the
amount of $k$. Besides, when $r$ tends to zero, the mentioned invariants are
given as
\begin{equation*}
\begin{array}{cl}
\left. \mathcal{K}\right\vert _{{r\rightarrow 0}^{+}}= & 192\,\left( {\frac{\mathbb{G}%
\left( k-1\right) }{L}}-\frac{1}{4\,l^{2}}\right) ^{2}-{\frac{320\,\mathbb{G}%
k\left( k-1\right) }{M{L}^{2}}\left( {\frac{\mathbb{G}\left( k-1\right) }{L}}%
-\frac{1}{4\,l^{2}}\right) }{r}^{2}+O\left( {r}^{4}\right), \\
\left. \mathcal{R}^{2}\right\vert _{{r\rightarrow 0}^{+}}= & 192\,\left( {%
\frac{\mathbb{G}\left( k-1\right) }{L}}-\frac{1}{4\,l^{2}}\right) ^{2}-{%
\frac{320\,\mathbb{G}k\left( k-1\right) }{M{L}^{2}}\left( {\frac{\mathbb{G}%
\left( k-1\right) }{L}}-\frac{1}{4\,l^{2}}\right) }{r}^{2}+O\left( {r}%
^{4}\right), \\
\left. \mathcal{R}\right\vert _{{r\rightarrow 0}^{+}}= & 24
\left({\frac{\mathbb{G}\left( k-1\right) }{L}}-\frac{1}{4\,
l^{2}}\right)-{\frac{20\,\mathbb{G}k\,\left(
k-1\right) }{M{L}^{2}}}{r}^{2}+O\left( {r}^{4}\right).%
\end{array}%
\end{equation*}
\begin{table}[tbp]
\caption{The values of curvature invariants for $k=4$, $M=0.17$,
$\mathbb{G}=l=1$, $L=3$} \label{tab1}\centering
\begin{tabular}{|c|c|c|c|}
\hline $r$ & $\mathcal{R}^2$ & $\mathcal{R}$ & ${\mathcal{K}}$ \\
\hline
0.0 & 108.00 & 18.00 & 108.00 \\
0.2 & 52.85 & 12.51 & 55.00 \\
0.4 & 6.09 & 2.55 & 17.82 \\
0.6 & 9.00 & -3.82 & 21.42 \\
0.8 & 14.45 & -6.10 & 20.64 \\
1.0 & 14.79 & -6.45 & 16.91 \\
1.2 & 13.85 & -6.40 & 14.46 \\
1.4 & 13.07 & -6.25 & 13.24 \\
1.6 & 12.60 & -6.14 & 12.64 \\
1.8 & 12.33 & -6.08 & 12.34 \\
2.0 & 12.19 & -6.04 & 12.19 \\ \hline
\end{tabular}%
\end{table}
\begin{figure}[!htb]
    \centering  {%
        \includegraphics[scale=0.3
        ]{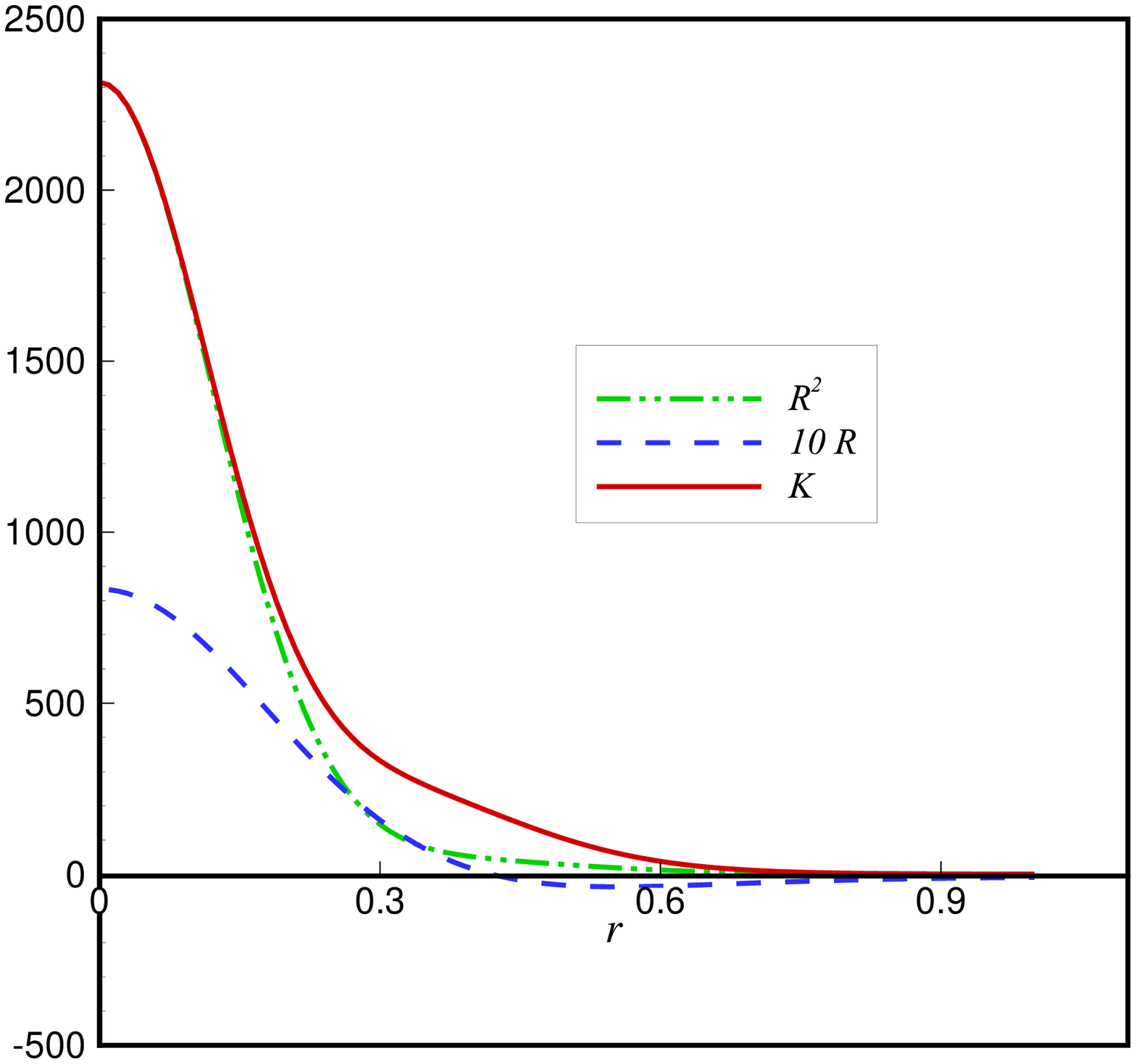}\label{fig8a}} \hspace*{.07cm}
    {\includegraphics[scale=0.3]{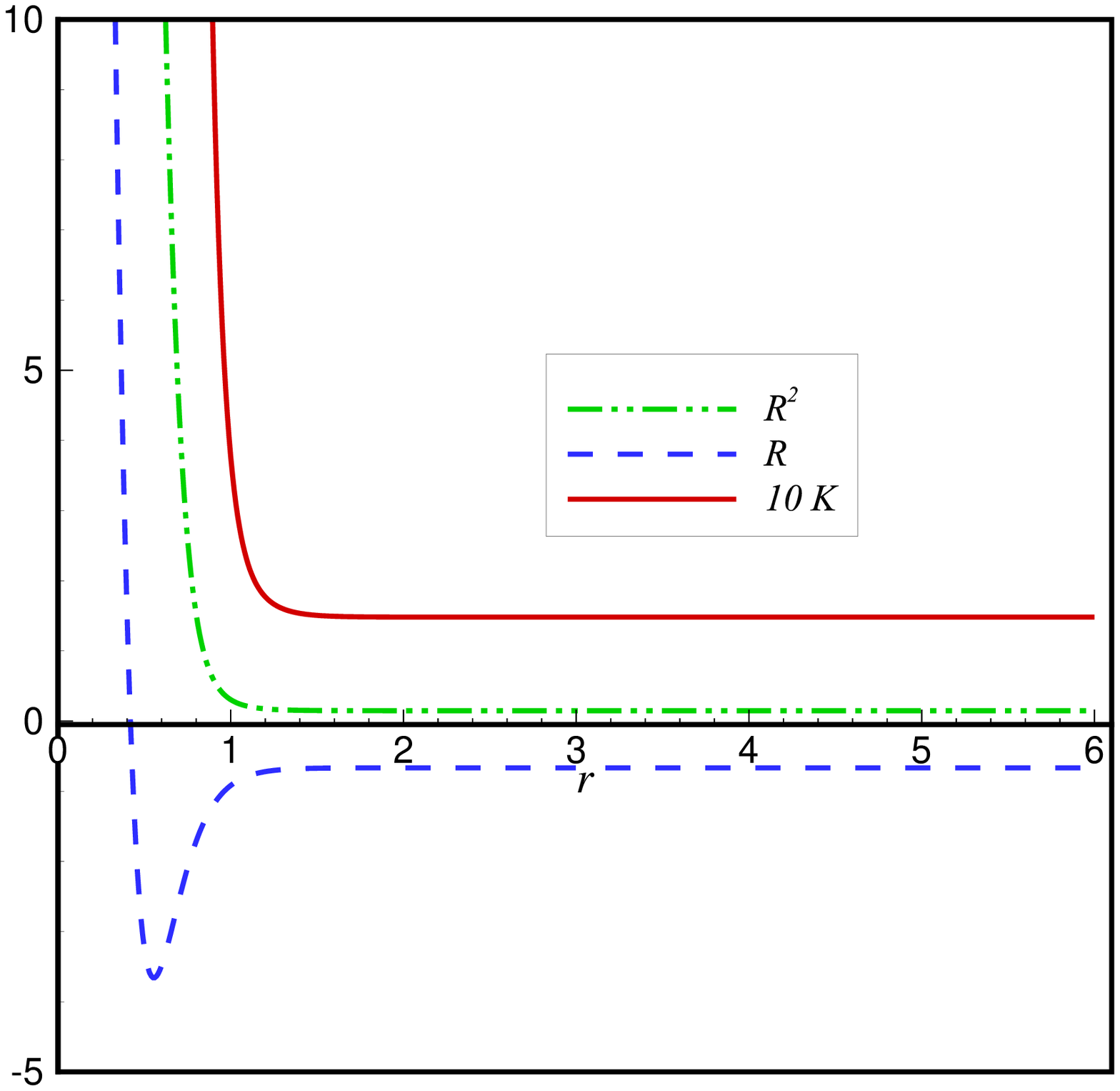}\label{fig8b}}
    \caption{Behavior of curvature invariants with respect to $r$
    for $k=8$, $ M=0.2$, $\mathbb{G}=1$, $L=2$ and $l=3$ (left and
    right panels are plotted with different scales).}\label{fig8}
\end{figure}

\begin{figure}[!htb]
    \includegraphics[scale=0.3]{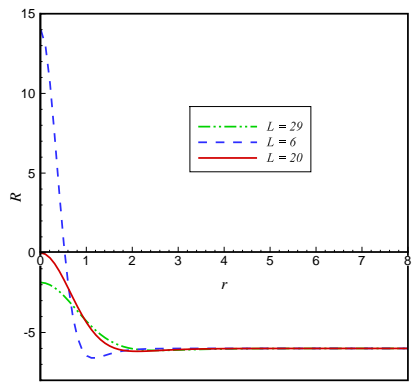}
    \caption{Behavior of the Ricci scalar with respect to $r$ for $k=6$,
    $\mathbb{G}=1$, $l=1$ and $L=6$(dS core), $L=20$ (flat core) and $L=29$ (AdS core).}
    \label{fig10}
\end{figure}


To be more clear, we have provided table \ref{tab1}. As it is
clear, none of the curvature invariants diverges neither at the
origin nor other values of $r$. Furthermore, Fig. \ref{fig8} shows
the behavior of curvature invariants for other classes of metric
parameters which confirms that they are free of divergencies as we
expect for the regular black holes.

Before proceeding, it is worthwhile to study the behavior of the
Ricci scalar by considering the structure of the black hole's
core. Investigating the behavior of this function near the origin
indicates that depending on the values of the metric parameters,
$\mathcal{R}$ can be zero, positive or negative at $r \rightarrow
0^{+}$. To make this point more clear, the behavior of the Ricci
scalar with respect to $r$ is sketched for different values of $L$
and fixed values of the other parameters in Fig. \ref{fig10}.
Since the asymptotic behavior of the solution is AdS, this
function for the regular black hole with dS core enjoys a root at
a finite radius and after having a local extremum, it tends to a
constant value $(-6/l^2)$ at infinity. However, the Ricci scalar
for the black hole whose core geometry is AdS (flat) is a smooth
function of $r$ and its value starts from a negative point (zero)
and tends to a constant value $(-6/l^2)$ at infinity. As a final
comment, it is notable that the core discussion based on the
behavior of curvature invariants near the origin is in agreement
with what we mentioned before after Eq. (\ref{f0}).

\begin{figure}[!htb]
\centering \subfigure[{\,$\mathbb{G}=L=l=1$, $k=2$}] {%
\includegraphics[scale=0.26
        ]{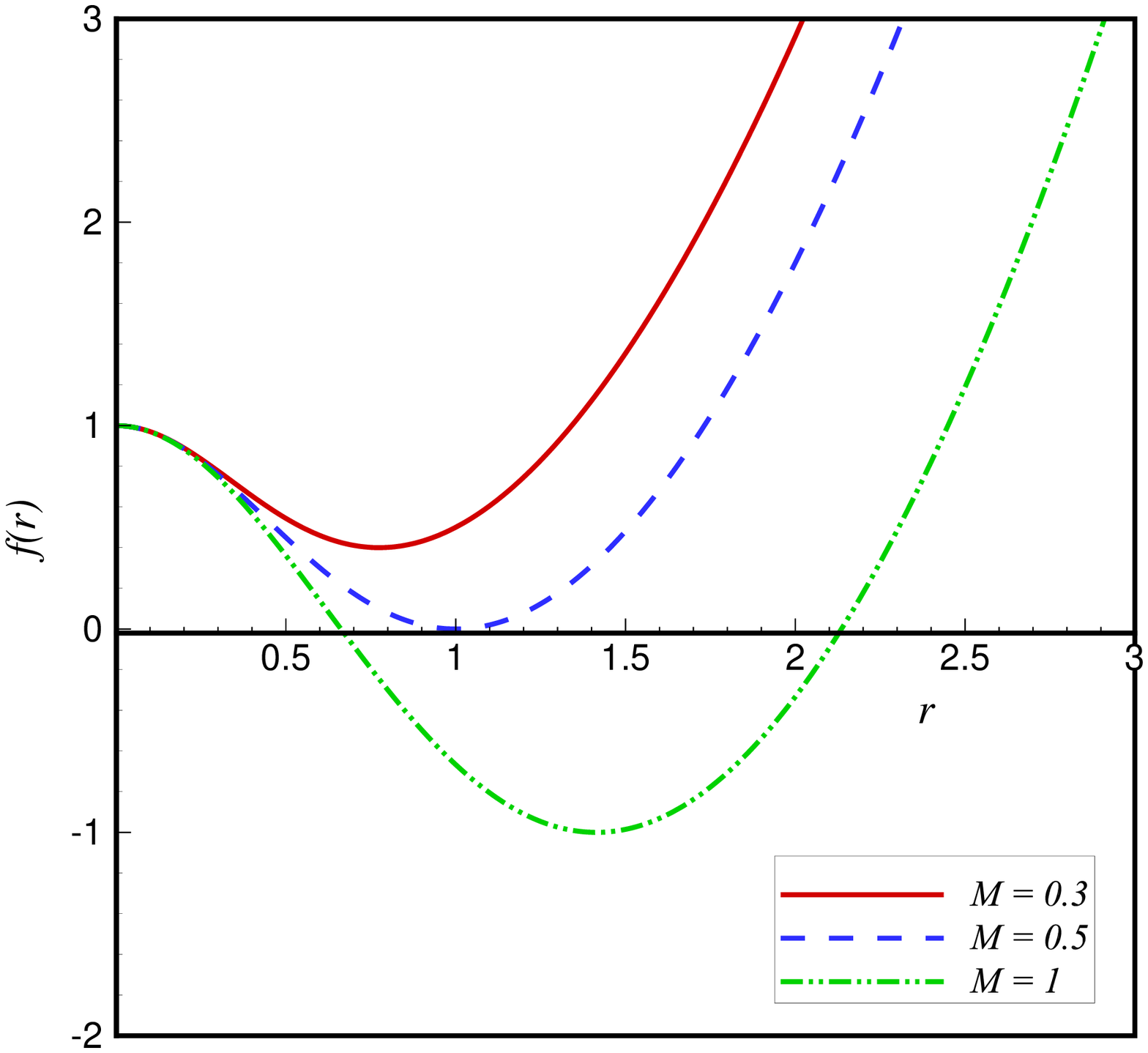}\label{f01}} \hspace*{.07cm}
\subfigure[{\,$\mathbb{G}=L=M=1$,
$k=2$}]{\includegraphics[scale=0.26]{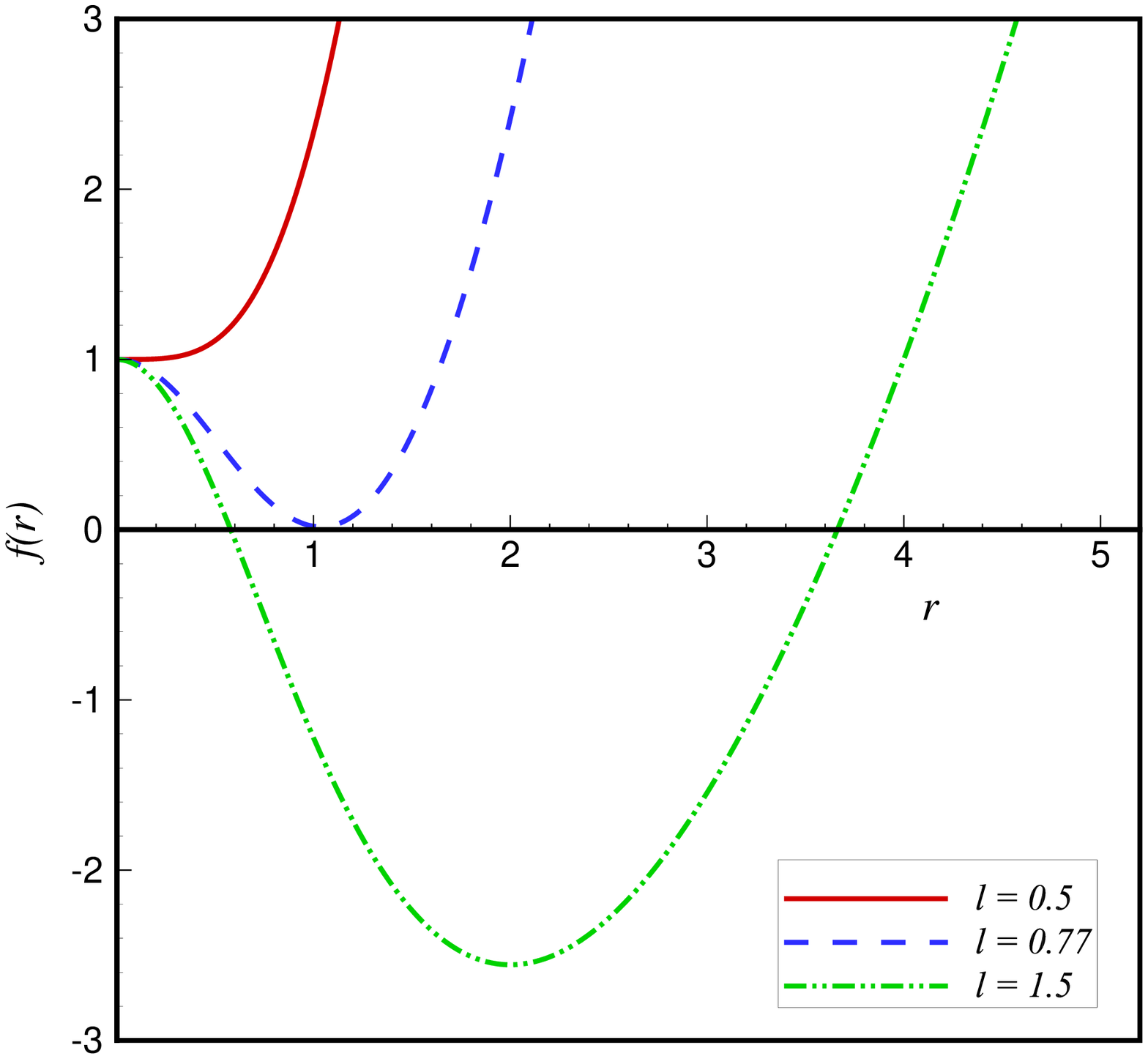}\label{f02}} \hspace*{.07cm}
\subfigure[{ \,$\mathbb{G}=L=M=1$,
$l=0.515$}]{\includegraphics[scale=0.26]{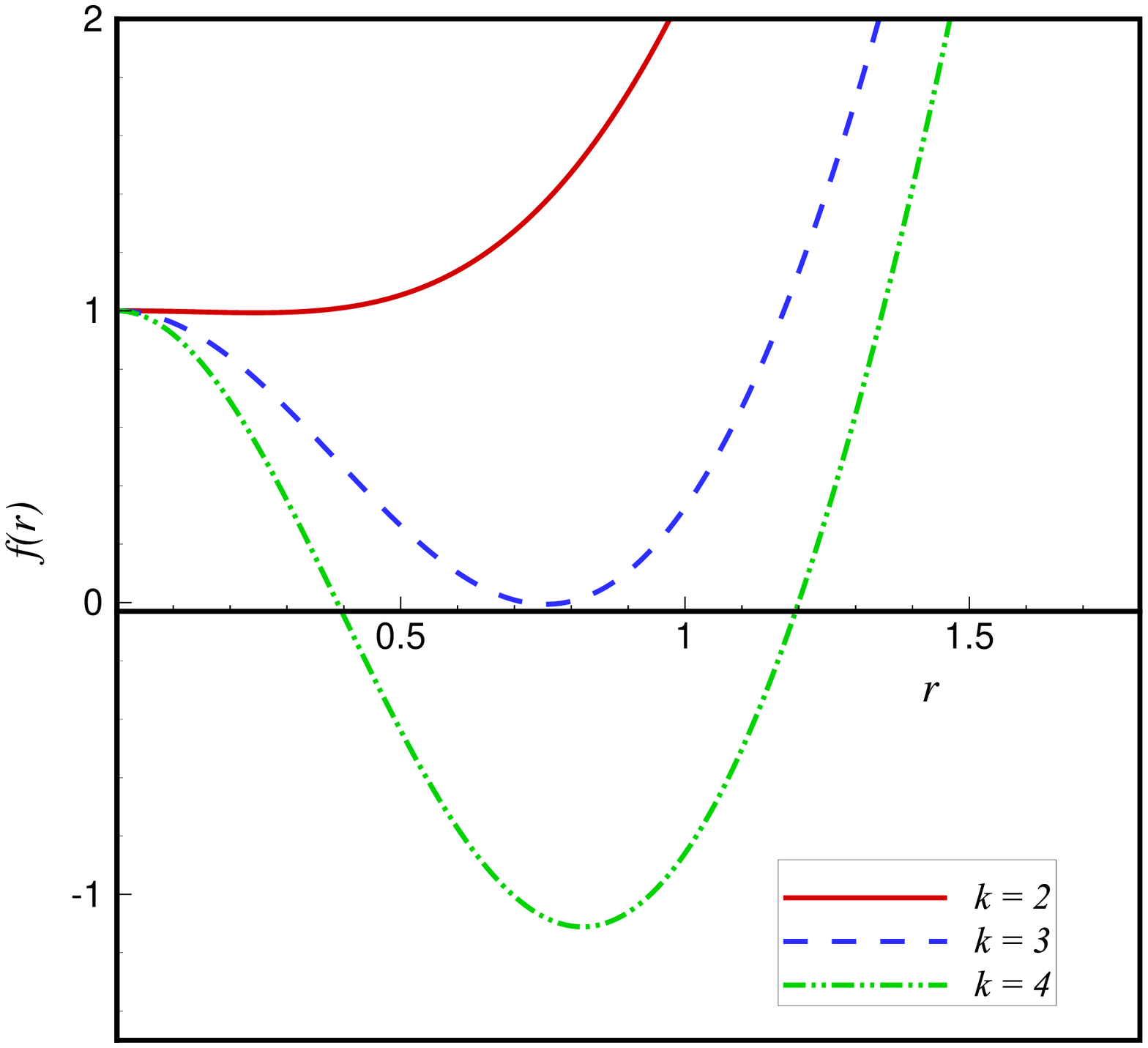}\label{f03}}
\caption{Behavior of $f(r)$ with respect to $r$}
\label{f}
\end{figure}

Now, we try to investigate other physical properties of the
solution by studying the behavior of the obtained solution and
looking for the horizons. In this regard, we have plotted the
function $f(r)$ versus $r$ for different model parameters in Fig.
\ref{f}. These figures show that, depending on the metric
parameters, this solution could represent a black hole with two
horizons or an extreme black hole with a degenerate root. In the
case of positive definite $f(r)$, we have a regular horizonless
spacetime which we are not interested in. Figure \ref{f01}
indicates by increasing the value of the mass parameter (and fixed
values of the other parameters), the number of horizons changes
from one to two. Also, considering Fig. \ref{f02} and Fig.
\ref{f03}, we find that by decreasing the values of parameters $l$
and $k$, two horizons merge to a degenerate one creating an
extremal solution.

To confirm our claims regarding the number of horizons, we calculate $f(r)$
derivative's roots. Since the number of function derivative's roots
represents the number of extrema of the function, if $f(r)$ has more than
one extrema, the number of horizons could be more than two. However, the
existence of just one extremum represents that the maximum number of horizon
will be two. In case of our solution, $f^{\prime }(r)$ has three roots as
\begin{eqnarray}
r\Big|_{f^{\prime }=0}=\left\{0,\,\pm\sqrt {2LM \left[ \left({\frac {L}{4\,%
\mathbb{G}\,{l}^{2} \left( k-1 \right) }} \right) ^{-{1/k}}-1 \right]}%
\right\},
\end{eqnarray}
where $f^{\prime }=\frac{\partial}{\partial r}f(r)$. Since there
is only one acceptable extremum (real and positive), it can be
ensured that the maximum number of horizons is two.\par Elementary
analysis regarding to roots of $f(r)$ reveals a special mass
\begin{equation}
M_{ext}=\frac{1}{8\,\mathbb{G}}\Bigg{\{} 1+{\frac {L}{4\,\mathbb{G} \left( k-1 \right) {l}^{
            2}} \left(  \left[ 1- \left({\frac {L}{4\,\mathbb{G} \left( k-1 \right) {l}^
            {2}}} \right) ^{-\frac{1}{k}} \right] k-1 \right) } \Bigg{\}} ^{-1}
\end{equation}
such that for $M$ larger than $M_{ext}$, $f(r)$ enjoys two simple
roots at $r=r_\pm$ and for $M=M_{ext}$ ($M<M_{ext}$), $f(r)$ has a
degenerate zero at $r=r_{ext}$ (no roots). Due to the complexity
of the equations, it is difficult to find the horizon radius of
extreme black hole for all values of $k$ parameters. However, in
the case of $k=2$, mass and horizon radius of the solution will be
obtained as follows
\begin{eqnarray}
&& M_{ext}\Big|_{k=2}=\frac{1}{8\,\mathbb{G}}\left[1+\frac{L}{4l^2\,\mathbb{G}}\left(1-4l\sqrt{{\mathbb{G}}{L^{-1}}}\right)\right]^{-1},\nonumber\\
&&\nonumber\\&&
r_{ext}\Big|_{k=2}=\sqrt{\frac{l}{2\sqrt{{\mathbb{G}}{L^{-1}}}-1}}.
\end{eqnarray}

\begin{figure}
    \centering   \subfigure[{$k=2$}]
    {   \includegraphics[scale=0.3
        ]{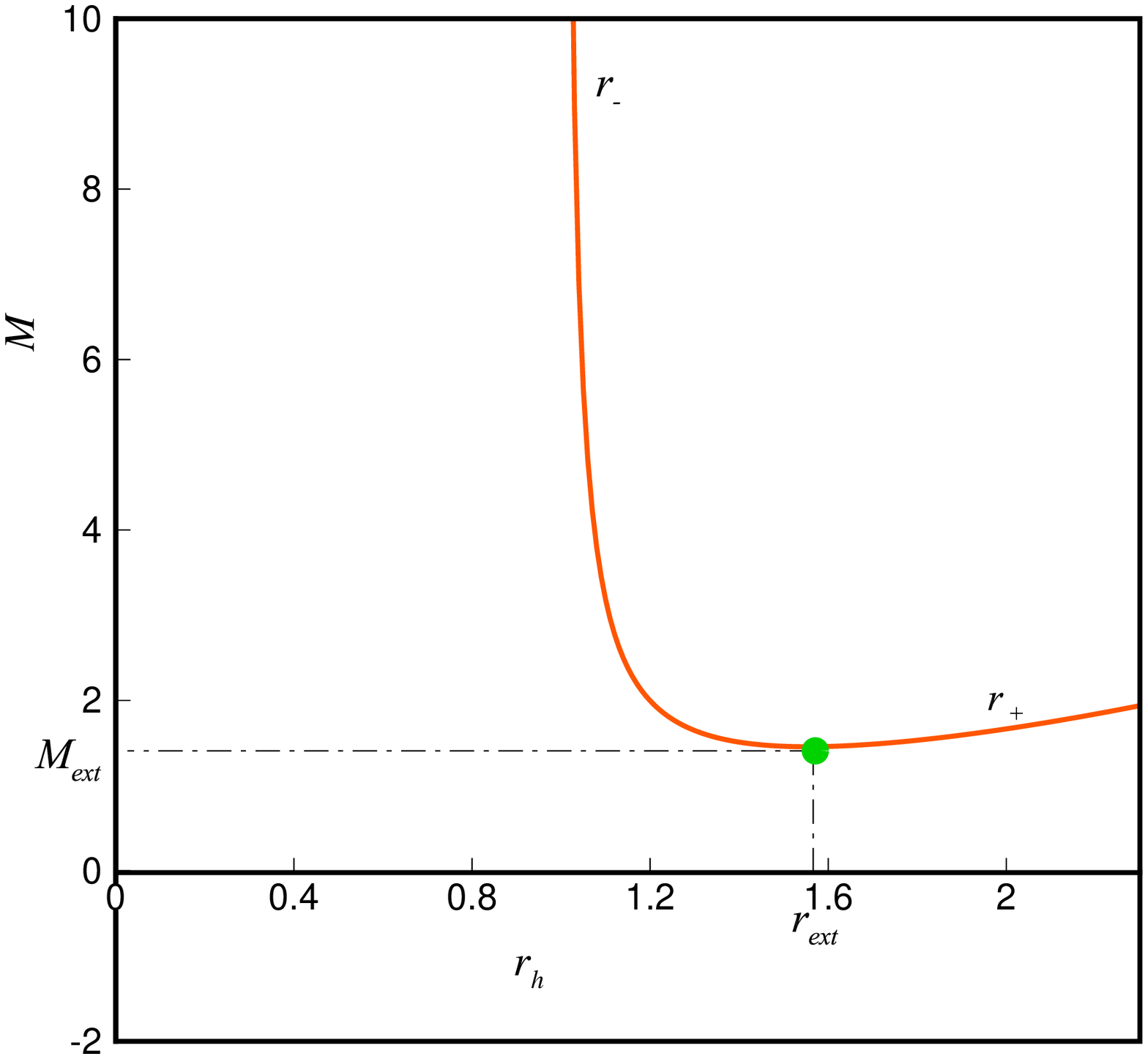}\label{mext1}} \hspace*{.07cm}
    \subfigure[{different $k$}] {\includegraphics[scale=0.3]{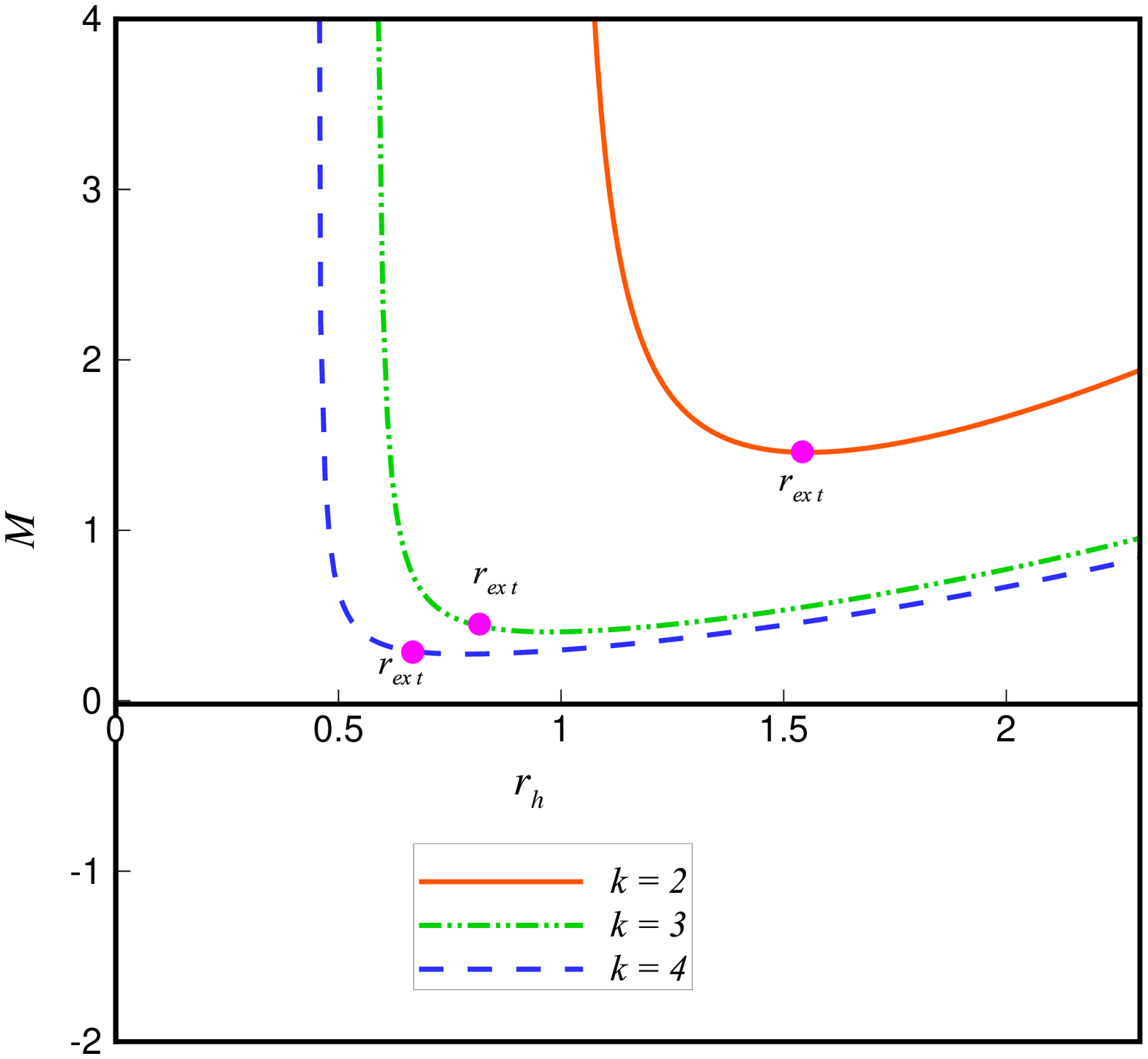}\label{mext2}}
    \caption{{Horizon mass-radius relation
            for $L=l=1$ and $\mathbb{G}=1$ }}\label{mext}
\end{figure}

Here, it is worth discussing the evolution of the mass parameter
$M$. To this end, we use Eq. \eqref{f metric} to derive the
relation between the mass parameter and the horizon radius which
for the case $k=2$ will be as follows
\begin{equation}
M(r_{h})|_{k=2}=\frac{r^2}{2\,L}\left(\frac{4\,\mathbb{G}\,{l}^{2}{r_{h}}^{
        2}}{L\left( {l}^{2}+{r_{h}}^{2} \right)}-1\right)^{-1},
\end{equation}
corresponding to the root of the equation $f(r_{h}) = 0$. To
better understand this function, the evolution of the mass
parameter as a function of the horizon radii is displayed in Fig
\ref{mext1}. As can be seen from this figure, there is a critical
value of the mass parameter $M_{ext}$, corresponding to the
minimum value on the curve. At this point where the inner ($r_-$)
and outer horizon ($r_+$) coincide, the solution meets the extreme
black hole condition. However, the proposed regular black hole
enjoys a pair of horizons when mass parameter exceeds critical
mass $M_{ext}$. Moreover, one can notice that
\begin{equation}\label{MassesAsFunction}
\frac{d M}{dr_{-}} \leq 0   \textrm{ ,\,\,\,\,\, } \frac{d M }{d r_{+}} \geq 0,
\end{equation}
implying the fact that decreasing the value of $r_+$ results in decreasing the mass parameter.
To investigate the effects of the parameter $k$ on the evolution
function of mass parameter $M(r_{h})$, we have provided Fig.
\ref{mext2} which shows that increasing the value of the $k$
parameter leads to decreasing the value of $M_{ext}$ and $r_{ext}$
and, therefore, it causes the diagram to be inclined towards the
origin.

\begin{figure}[!htb]
\includegraphics[scale=0.4]{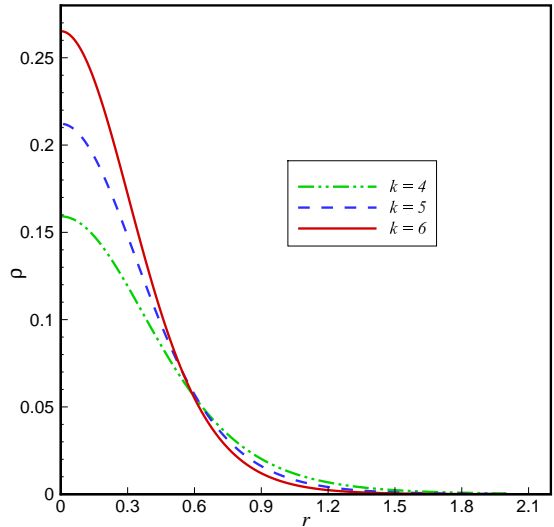}
\caption{Behavior of $\protect\rho $ with respect to $r$ for $M=0.2$ and $L=3
$}
\label{fig9}
\end{figure}

{In the following, we study some other properties of the proposed
energy density.} Regarding the behavior of the energy density with
respect to $r$, we calculate its derivative as
\begin{eqnarray}
\frac{d \rho}{d r}=-\,\frac { \left( k-1 \right) kr}{2\pi \,{L}%
^{2}M} \left( 1+\,{\frac {{r}^{2}}{2LM}} \right) ^{-k-1},
\end{eqnarray}
where shows that due to the positive values of $L$, the suggested function
of energy density is strictly decreasing, having the maximum value at the
origin and zero at infinity (see Fig. \ref{fig9}).

According to Fig. \ref{fig9}, we find that {, as we expect,}
$\rho$ is finite at the origin. {It is discussed the reason for
the finiteness of the energy density at the origin in the appendix
\ref{appendix1}.} Also, increasing $k$ (when other metric
parameters are fixed) can increase the amount of $\rho$ at $r=0$
but keeps it
finite. An important note is that $k$ is a finite parameter since when $k\to \infty$%
, the energy density vanishes ($\rho\to 0$). To investigate the behavior of
the energy density on the event horizon, the value of $\rho$ should be
calculated as a function of $r_{+}$ which will take the following form
\begin{equation}  \label{rhoplus}
\rho_{+}\,=\,{\frac {\left( k-1 \right)\,M }{ \pi \,r_{+}^2} \big( 1-\,\eta %
\big) \left[1- \left( 1-\,\eta \right) ^{\frac{1}{k-1}} \right] },
\end{equation}
where
\begin{equation}
\eta={\frac {1}{8\,\mathbb{G}\,M} \left( 1+{\frac {r_{+}^{2}}{{l}^{2 }}}
\right) }.
\end{equation}
It should be mentioned that to arrive the above result we have used the
following relation
\begin{equation}  \label{L}
L=\frac{r_{+}^{2}}{2M}\left\{\left({1-\eta}\right)^{\left( \frac{1}{1-k}%
\right)}-1 \right\}^{-1},
\end{equation}
to eliminate $L$, which comes from the condition of the event horizon, i.e. $%
f(r=r_{+})=0$.
\begin{figure}[!htb]
\begin{center}
\includegraphics[scale=0.3]{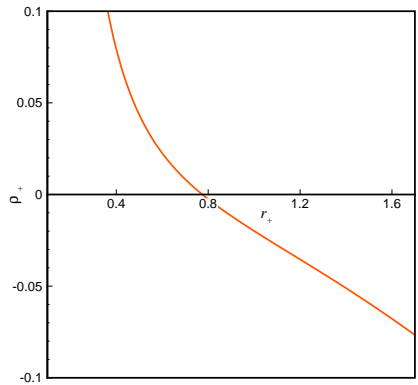}
\end{center}
\caption{Behavior of $\protect\rho_{+}$ with respect to $r_+$}
\label{fig6}
\end{figure}
To study more closely, we have provided Fig. \ref{fig6} in which the
behavior of $\rho_{+}$ in terms of $r_{+}$ is depicted. According to these
figures (and also Eq. \eqref{rhoplus}), the energy density will be a
positive real parameter if
\begin{equation}
r_{+}^2<r^{2}_{\max}=(8M\mathbb{G}-1)l^2,  \label{upper}
\end{equation}
which means that there is an upper limit for the event horizon radius.
Moreover, a lower limit will be placed on the mass parameter, i.e. $M>\frac{1%
}{8\,\mathbb{G}}$, since the event horizon must be a real
parameter.

One of the interesting results of this paper is obtaining an upper
limit on the event horizon radius as introduced in Eq.
(\ref{upper}). The existence of the upper limit for the event
horizon is in direct contradiction to classical black holes such
as Schwarzschild and Reissner-Nordstr\"{o}m whose radius of their
event horizon is allowed to go to infinity. However, it does not
make sense that the radius of the event horizon tends to infinity.
Since there are some astrophysical black holes with an upper limit
reported on their mass \cite{Natarajan:2008ks}, it is expected
that there will be an upper limit for their radius of event
horizon, which is consistent with our result.



\section{Thermodynamics \label{thermo}}

Black hole thermodynamics in AdS space is interesting from the
AdS/CFT correspondence point of view that suggests the existence
of a holographic duality between quantum gravity on AdS space and
a certain Euclidean conformal field theory on its spacelike
boundary. Besides, black hole remnant that may give a solution to
the information paradox, can be considered in the context of black
hole thermodynamics. In what follows, we study the thermodynamic
feature of the obtained solution.

\subsection{Conserved charge of the solution}
In this subsection, we compute a conserved charge for our solution
employing the method describing in \cite{Aoki:2020prb}. According
to \cite{Aoki:2020prb}, for any classical or quantum field theory
on a general curved spacetime, the following quantity
\begin{equation}
    Q(t) := \int_{\Sigma_t} d^{d-1}\vec x\,  \sqrt{\vert g\vert} J^0 (t,\vec x),
\end{equation}
is conserved under the given time evolution, where $\Sigma_t$ is a
hypersurface or a time slice of the spacetime $\Sigma$ at an
arbitrarily fixed time $t$, $d$ is the dimension of the spacetime
$\Sigma$, and $g$ denotes the determinant of $g_{\mu\nu}$.
Moreover, $ J^0$ is the zero component of a covariantly conserved
current $J^\mu$, $\nabla_\mu J^\mu = 0$, where $\nabla_\mu$ is the
covariant derivative for the metric $g_{\mu\nu}$. For a
gravitational system with a Killing vector $\xi$ and a given
energy-momentum tensor $T^\mu{}_\nu$, the covariantly conserved
current can be constructed as follows
\begin{equation}
    J^\mu = T^\mu{}_\nu \xi^\nu.
\end{equation}

One can easily prove the covariantly conservation of this current
by using $\nabla_\mu T^\mu{}_\nu =0$ and $\nabla_{\mu}
\xi_{\nu}+\nabla_{\nu} \xi_{\mu} = 0$. Therefore, the defined
conserved charge will be a Noether charge corresponding to global
symmetry of the system.\par If $\xi^\mu$ is a Killing vector
associated with the time translation, the conserved charge will be
the total energy of the system
\begin{equation}
E = \int_{\Sigma_t} d^{d-1}\vec x \sqrt{\vert g\vert} T^0{}_\mu \xi^\mu,
\end{equation}
which will be in agreement with the standard definition of the
energy in the flat background with $\xi^\mu=-\delta^\mu_0$.\par
Following this method and choosing $\xi^\mu=-\delta^\mu_0$, the
total energy of the black hole corresponding to our model in
$(2+1)-$ dimensions will be as follows
\begin{equation}
E=-2\pi \int^\infty_0 dx x T^0_0= 2\pi\,\lim_{r \to \infty}
\int^r_0 dx x \rho(x)= \lim_{r \to \infty} m(r)=M,
\end{equation}
 which confirms that the mass parameter M represents the total energy of the
 black hole.

\subsection{First law of thermodynamics}

Here, in order to study thermodynamic properties of the obtained regular
black hole, we begin with calculating some thermodynamic quantities. As the
first step, we focus on the entropy of the black hole which is equal to a
quarter of the event horizon area since we are working in Einstein gravity
\cite{Bekenstein:1973ur, Hawking:1998jf}%
\begin{equation}
S=\left. \frac{1}{4}\int\limits_{0}^{2\pi }\sqrt{g_{\phi \phi }}d\phi
\right\vert _{r=r_{+}}=\frac{\pi r_{+}}{2}.  \label{entropy}
\end{equation}%
To investigate physical properties, we should determine temperature $T$ as
the next step. One of the conventional methods of calculating $T$ is using
the surface gravity ($\kappa $) interpretation and its relation to the
Hawking temperature as%
\begin{eqnarray}
T_{H} =\frac{\kappa }{2}=\left. \frac{f^{\prime }(r)}{4\pi }\right\vert
_{r=r_{+}}=\,{\frac{r_{+}}{2\pi \,{l}^{2}}}-{\frac{4\,\left( k-1\right)
\mathbb{G}\,M}{\pi \,r_{+}}}\left( {1-\,\eta }\right) {\left[ 1-\left(
1-\,\eta \right) ^{\frac{1}{k-1}}\right] },  \label{tem}
\end{eqnarray}%
where we have used \eqref{L} to eliminate $L$.

Taking the obtained entropy and temperature into account, we are
in a position to examine the first law of thermodynamics. However,
the first law of thermodynamics needs to modify for regular black
holes due to the inconsistency between Bekenstein-Hawking area law
and the conventional first law of black hole thermodynamics
arising from the inclusion of the matter fields in the
energy-momentum tensor \cite{Ma:2014qma}. Before proceeding and
finding a structure of the first law of thermodynamics for regular
black holes, it should be mentioned a point related to the
dependence of the mass function on the mass parameter $M$. In
fact, for spaces in which the asymptotic region is a proper limit,
$m(r)$ should tend to a constant $M$ which is in direct proportion
of the mass of the solution \cite{Aros:2019quj}. Therefore, $m(r)$
should be promoted to a function $m(r,M)$ such that $m(r,M)|_{M=0}
= 0$ \cite{Estrada:2019qsu}. Moreover, $m(r,M)$ should be an
increasing function with respect to the mass parameter. Thus,
    \begin{equation}
    \frac{\partial m(r,M)}{\partial M} > 0,
    \end{equation}
for arbitrary values of $r$.

With the mentioned point in mind, we try to obtain the corrected
form of the first law of thermodynamics for regular black holes by
variation of the function $f(r,M)$ with respect to its parameters,
i.e. $\delta f(r,M)|_{r=r_+}=0$. However, since the transformation
would be mapping a black hole into another black hole in the space
of solutions, the function $f(r_+,M)$ must still vanishes under
any transformation of the parameters. Therefore, $f(r_+,M)=0$ and
$\delta f(r,M)|_{r=r_+}=0$ are to be considered as constraints on
the evolution along the space of parameters \cite{Estrada:2019cig,
Estrada:2020tbz}. Thus, from the variation of the function
$f(r,M)$ we get
    \begin{equation}
    0 = (\frac{\partial f}{\partial r} dr)|_{r=r_+} + \frac{\partial f}{\partial M} dM.
    \end{equation}
    After some manipulations, one can find that the first law takes the following form
\begin{equation}
du=TdS,  \label{first law}
\end{equation}%
where
\begin{equation*}
du=\mathbb{G}\frac{\partial m(r_{+}, M)}{\partial M}\,dM.
\end{equation*}
It is worth mentioning that both terms, $du$ and $dS$, are local
variables defined at the horizon. This is while the modification
of the first law for regular black holes have also been
investigated by the inclusion of an extra factor corresponding to
an integration of the radial coordinate up to infinity
\cite{Ma:2014qma}. \par Using \eqref{entropy} and \eqref{tem}
along with \eqref{first law}, we checked the modified first law of
thermodynamics for regular black holes in the case of our model
and one can confirm this law is satisfied.


\subsection{Thermal stability\label{stability}%
} This subsection is devoted to the analysis of the black hole
thermal stability making use of the canonical and grand canonical
ensembles. Using the canonical ensemble method, one can explore
the local stability and phase transition points of the black holes
with regard to the sign/divergence points of the heat capacity.
Global stability of the black holes, on the other hand, can be
examined in the grand canonical ensemble by regarding the Gibbs
free energy of the black holes. It is worth mentioning that the
difference between the two types of stabilities lies in the fact
that in global stability, a system in equilibrium with a
thermodynamic reservoir is allowed to exchange energy with the
reservoir while local stability is related to how the
thermodynamical system responds to small variations of the
thermodynamic parameters. In what follows we study the local and
global stabilities of our proposed black hole in the canonical and
grand canonical ensembles.

\subsubsection{Local thermal stability in canonical ensemble }

For the purpose of investigating the black hole local stability,
we need to calculate the black hole heat capacity. In the
canonical ensemble, the positivity of the heat capacity guarantees
thermal stability of the solutions.{ It is notable that} the
temperature should be positive at the same time to ensure physical
solutions.

In our case study, there is an additional condition that is
related to the positivity of the parameter $L$ or equivalently the
positivity of the energy density. As mentioned earlier, positive
energy density impose an upper limit on the radius of the event
horizon. It is known that the heat capacity is defined as
\begin{equation}
C=T\,\frac{dS}{dT},  \label{C1}
\end{equation}%
where for the obtained solution, it gets the following form%
\begin{equation}
C=\frac{4\pi \,r_{+}\,(k-1)\mathbb{G}\,M}{\,B(r_{+})}\left[ 2-\left( 1-\eta \right)
^{\frac{k}{k-1}}-8\,\mathbb{G}\,M\,k\,\eta \right],  \label{heat}
\end{equation}%
where%
\begin{equation*}
B (r_{+})\,=\,\left[ 8\,\mathbb{G}\,M\left( k+k\,\eta -1\right) -2k+1\right] \left[
\left( 1-\eta \right) ^{\frac{1}{k-1}}-1\right] +\left( {8\,\mathbb{G}%
\,M\,\eta -1}\right) \left( {1-\eta }\right) {^{\frac{1}{k-1}}}.
\end{equation*}%
First, we consider the special case $k=2$ to compare our result with \cite%
{Estrada:2020tbz}. It is notable that for the case $k=2$ all relations are
simpler which is another reason to focus on this value. For this special
case, the heat capacity reduces to%
\begin{equation}
\left. C\right\vert {_{k=2}}=\frac{\pi \,r_{+}\left( {{\frac{8\,\mathbb{G}%
\,M\,\eta -1}{8\,\mathbb{G}\,M-1}}\left( {1+8\,\mathbb{G}\,M\,\eta }\right)
-1}\right) }{2\left( {1+{\ {\frac{8\,\mathbb{G}\,M\,\eta -1}{8\,\mathbb{G}%
\,M-1}}}\left( {24\,\mathbb{G}\,M\,\eta -1}\right) }\right) }{\ }.
\label{heatk2}
\end{equation}%
To investigate thermal stability more precisely, we have provided Figs. \ref%
{hc1}-\ref{fig3}.
\begin{figure}[!htb]
\begin{center}
\includegraphics[scale=0.3]{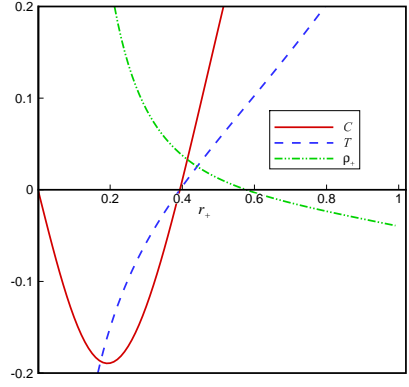}
\end{center}
\caption{Behavior of $C$, $T$ and $\protect\rho_{+}$ respect to $r_{+} $ for
$k=2$, $l=\mathbb{G}=1$ and $M=1/6$}
\label{hc1}
\end{figure}
The behavior of temperature, heat capacity and energy density in terms of $%
r_{+}$ are depicted in these figures. Regarding the temperature and the heat
capacity, we find similar behavior to that authors have done in Ref. \cite%
{Estrada:2020tbz}. However, we get more new results which we will discuss in
what follows.

Figure \ref{hc1} shows that all three functions are positive in the interval
${{r}_{\min }}<r_{+}<{r}_{\max }$ , where $r_{\min }$ is the radius in which
temperature (heat capacity) is starting to get the positive values. Besides,
for $r_{+}>r_{\max }$ the energy density will be negative which is
unphysical. Therefore, if the radius of the event horizon is in the
mentioned interval, the black hole is thermally stable and meets the
necessary criteria for viable solutions.
\begin{figure}[tbh]
\centering \subfigure[\, $C$ versus
$r_{+}$]{\includegraphics[scale=0.25]{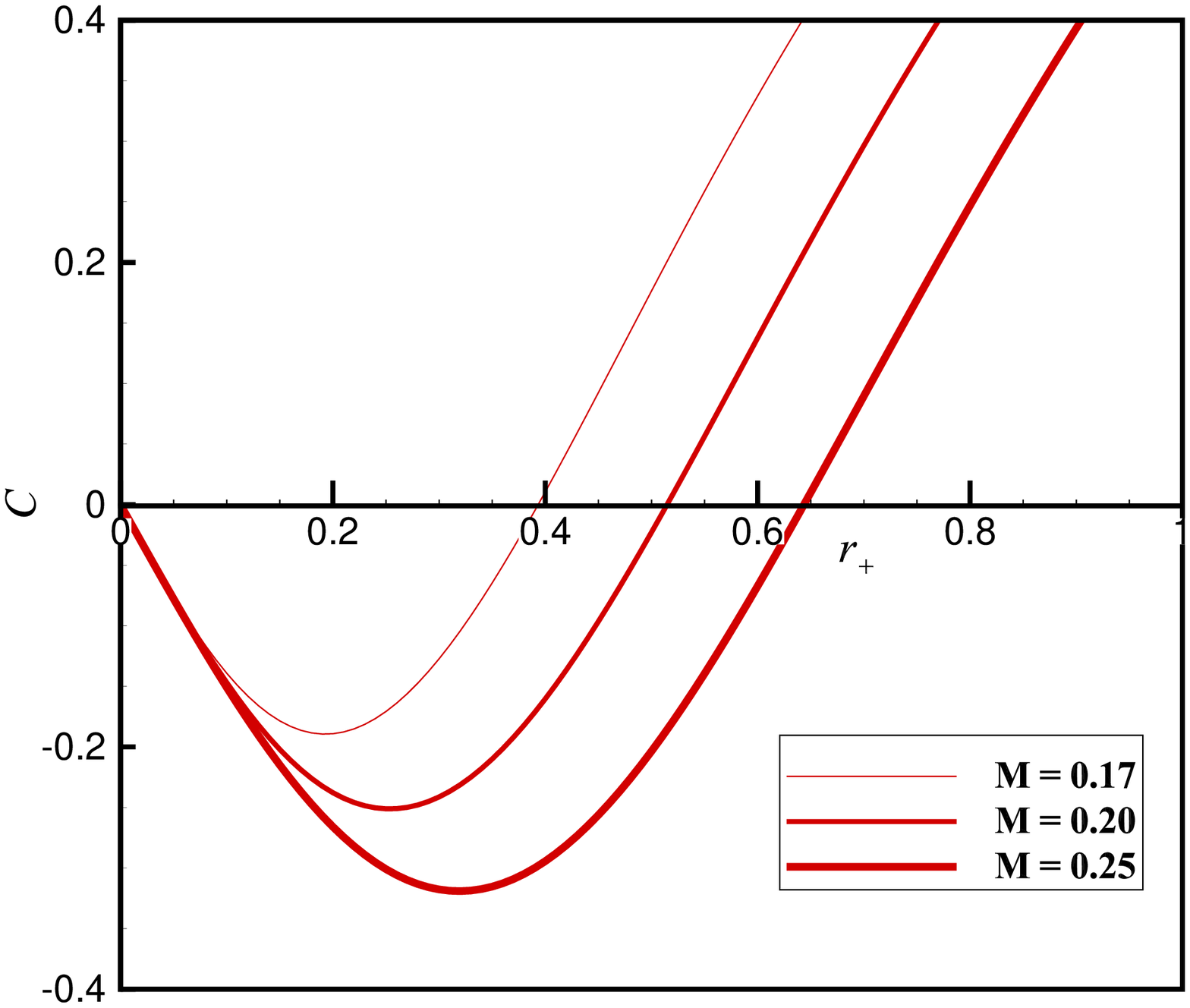}\label{fig2a}}
\hspace*{.1cm} \subfigure[\, $T$ versus $r_{+}$
]{\includegraphics[scale=0.25]{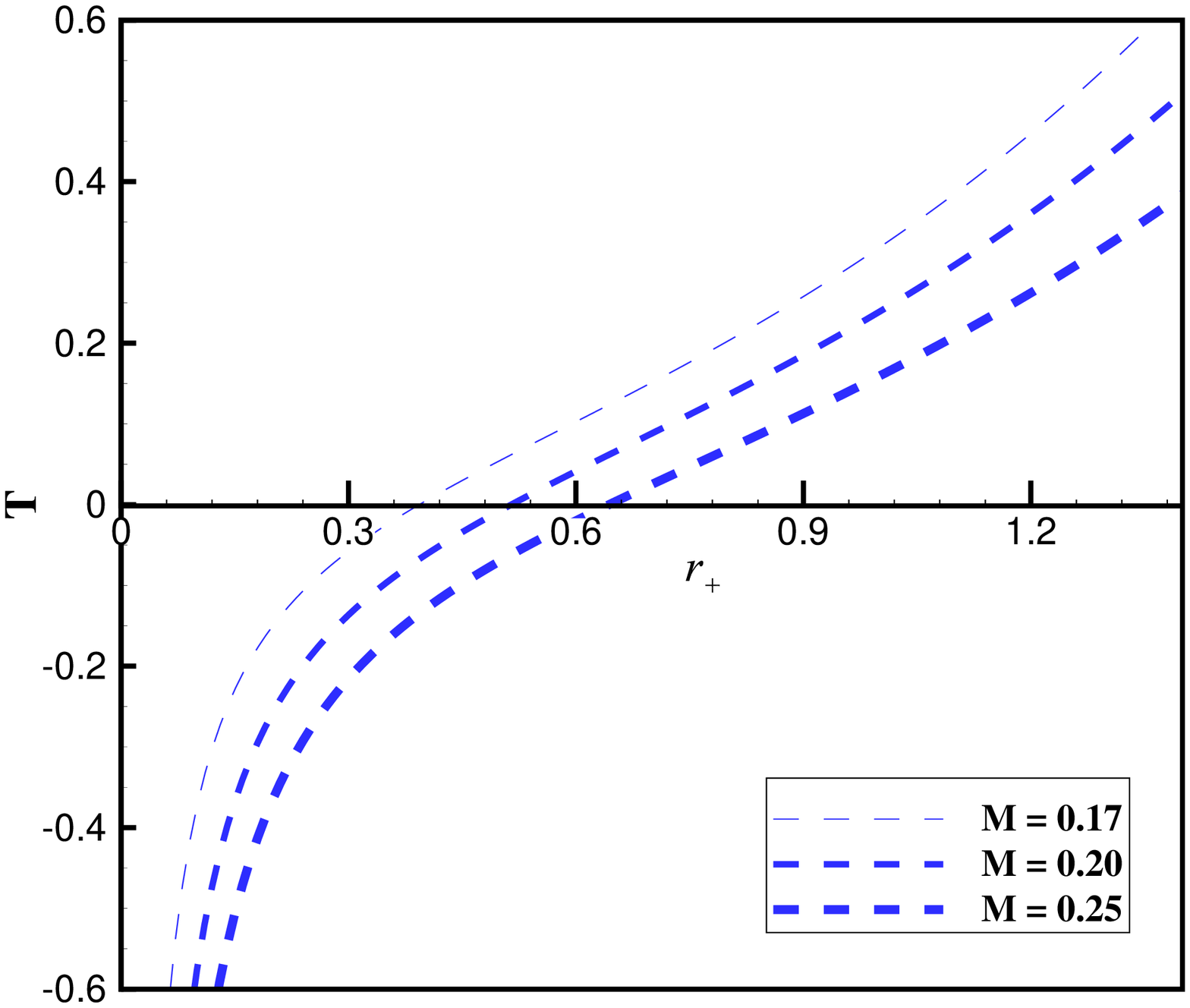}\label{fig2b}}
\hspace*{.1cm} \subfigure[\,$\rho_{+}$ versus $r_{+}$
]{\includegraphics[scale=0.25]{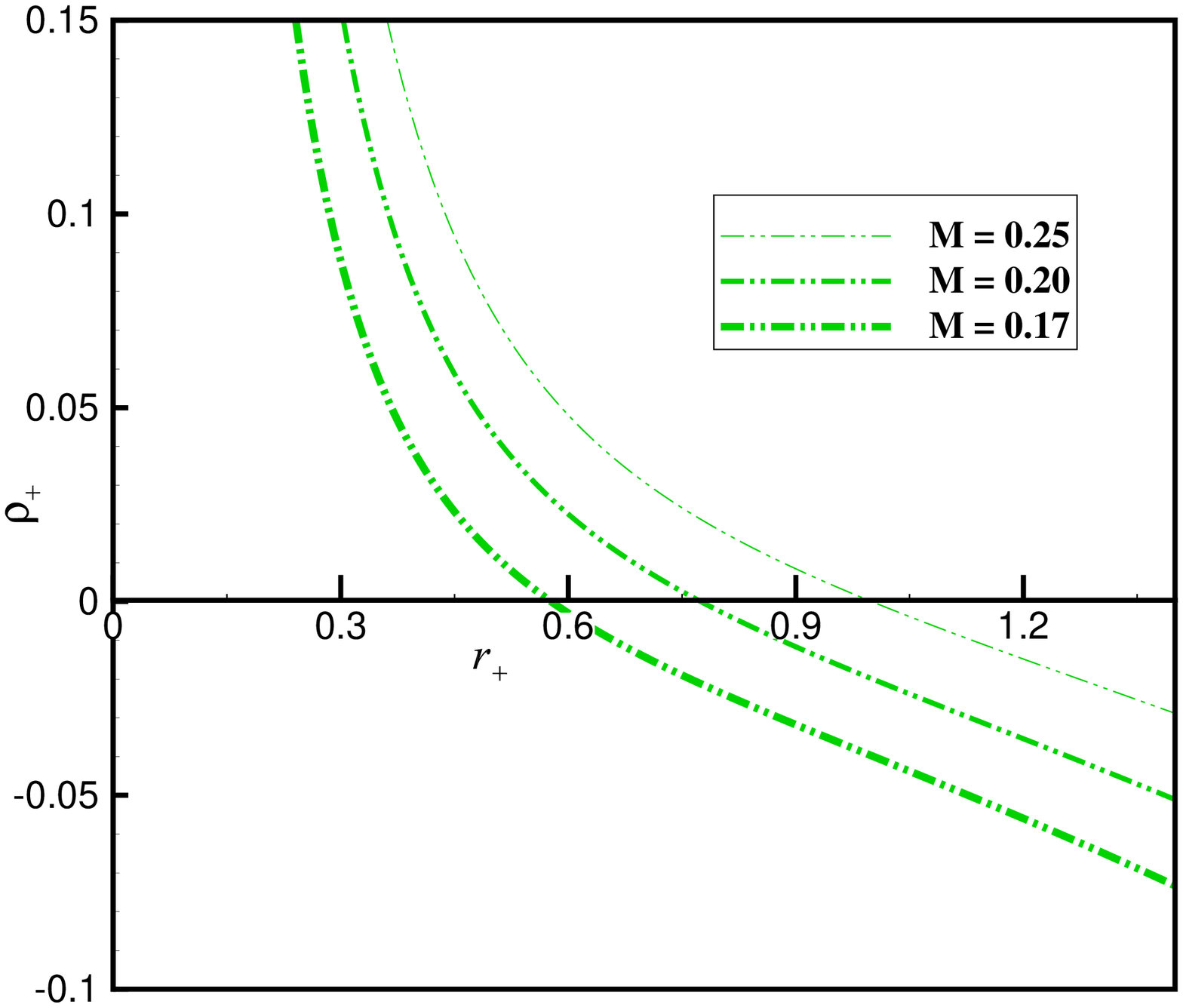}\label{fig2c}}
\caption{Behavior of $C$, $T$ and $\protect\rho _{+}$ respect to
$r_{+}$ for $\mathbb{G}=l=1$ and $k=2$} \label{fig2}
\end{figure}
This admissible domain can be altered by changing the metric
parameters. According to Fig. \ref{fig2}, increasing the values of
the mass parameter leads to increase the values of both
${{r}_{\min }}$ and ${r}_{\max }$ and makes the admissible domain
larger.
\begin{figure}[tbh]
\centering
\subfigure[\, $C$ versus
$r_+$]{\includegraphics[scale=0.25]{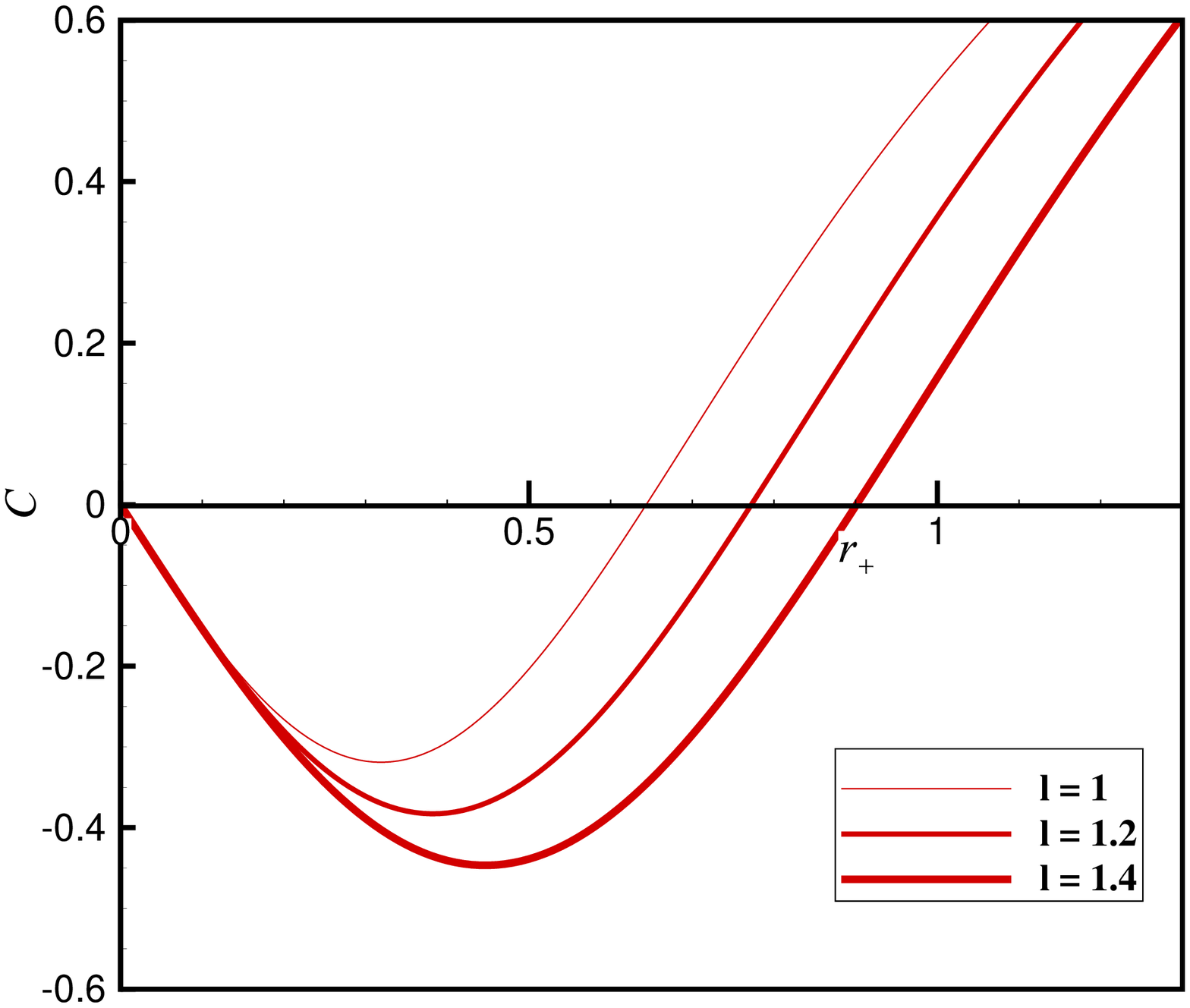}\label{fig3a}} \hspace*{.1cm}
\subfigure[\, $T$  versus $r_{+}$
]{\includegraphics[scale=0.25]{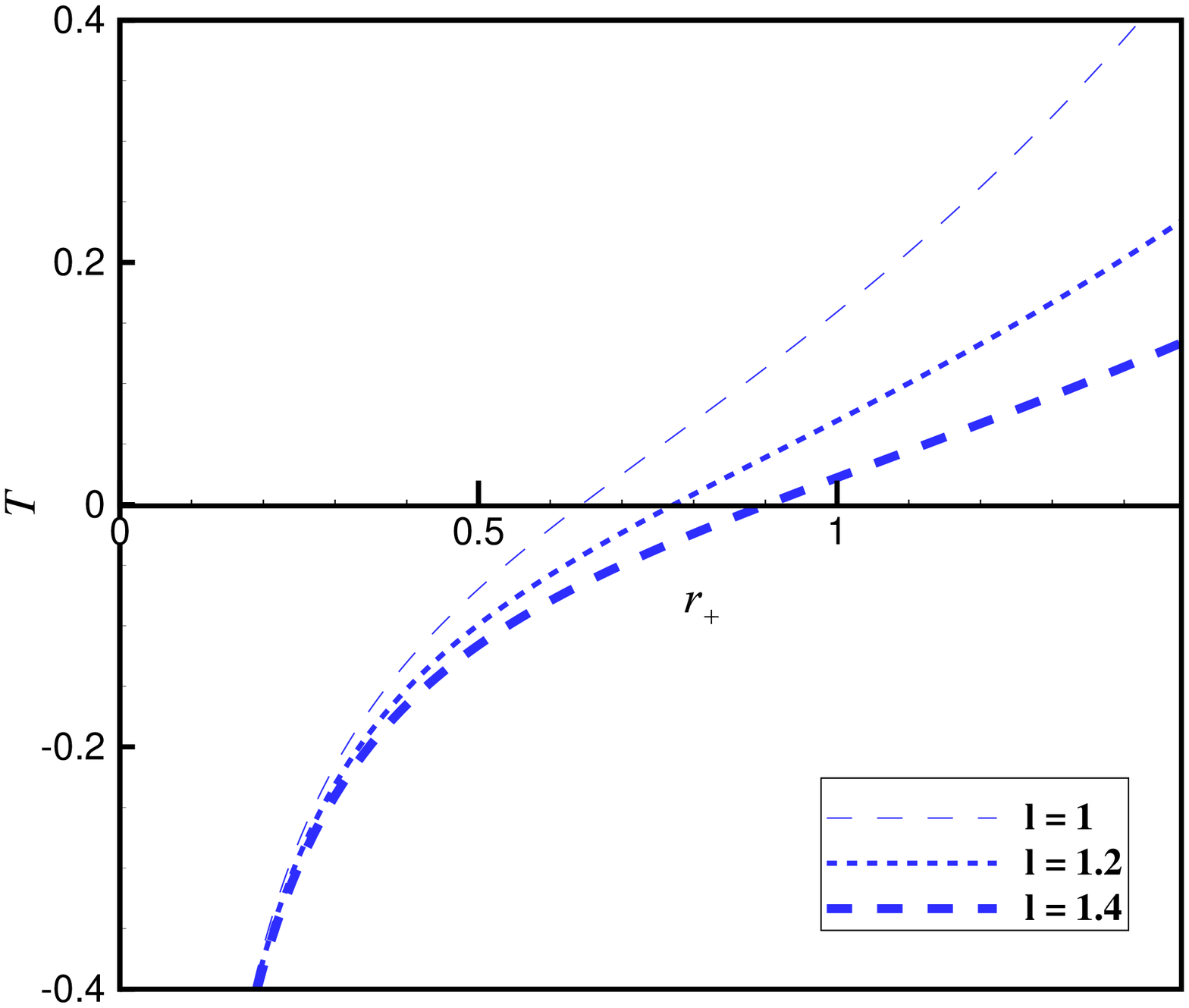}\label{fig3b}} \hspace*{.1cm}
\subfigure[\,$\rho_{+}$ versus $r_+$
]{\includegraphics[scale=0.25]{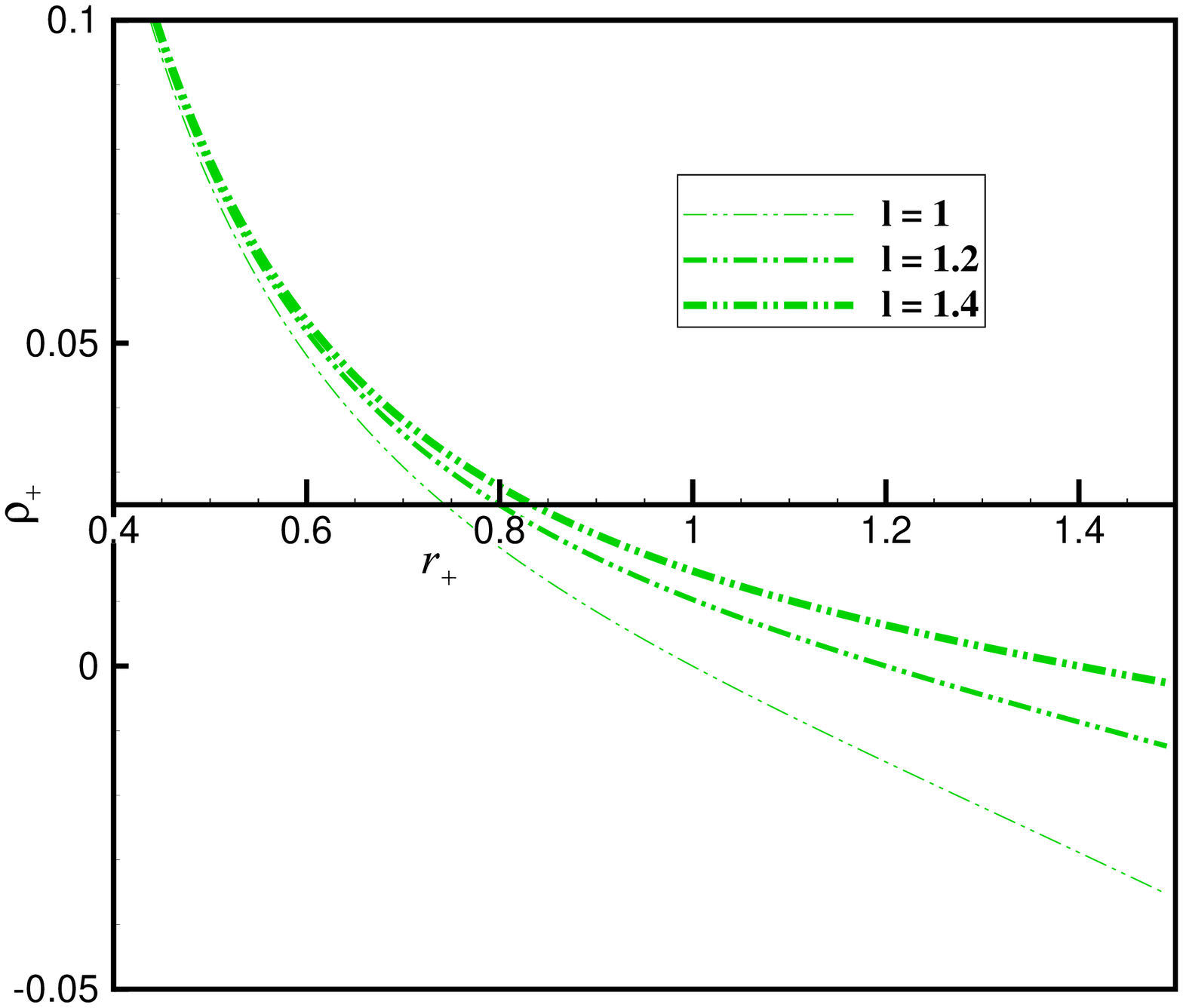}\label{fig3c}}
\caption{Behavior of $C$, $T$ and $\protect\rho _{+}$ respect to $r_{+}$ for
$\mathbb{G}=1$, $k=2$ and $M=0.25$}
\label{fig3}
\end{figure}
In addition, Fig. \ref{fig3} represents that increasing $l$ leads
to increasing both special limits ( ${r}_{\min }$ and ${{r}_{\max
}}$) and hence, the variation of this parameter can also change
the stability interval. To elucidate, the results for a range of
changes of $M$ and $l$ are displayed in table \ref{table}. As one
can see from this table, by increasing the values of $M$ (left
panel) and $l$ (middle table), the admissible domain increases.

\begin{figure}[!htb]
\centering
\subfigure[\, $C$ versus
$r_{+}$]{\includegraphics[scale=0.25]{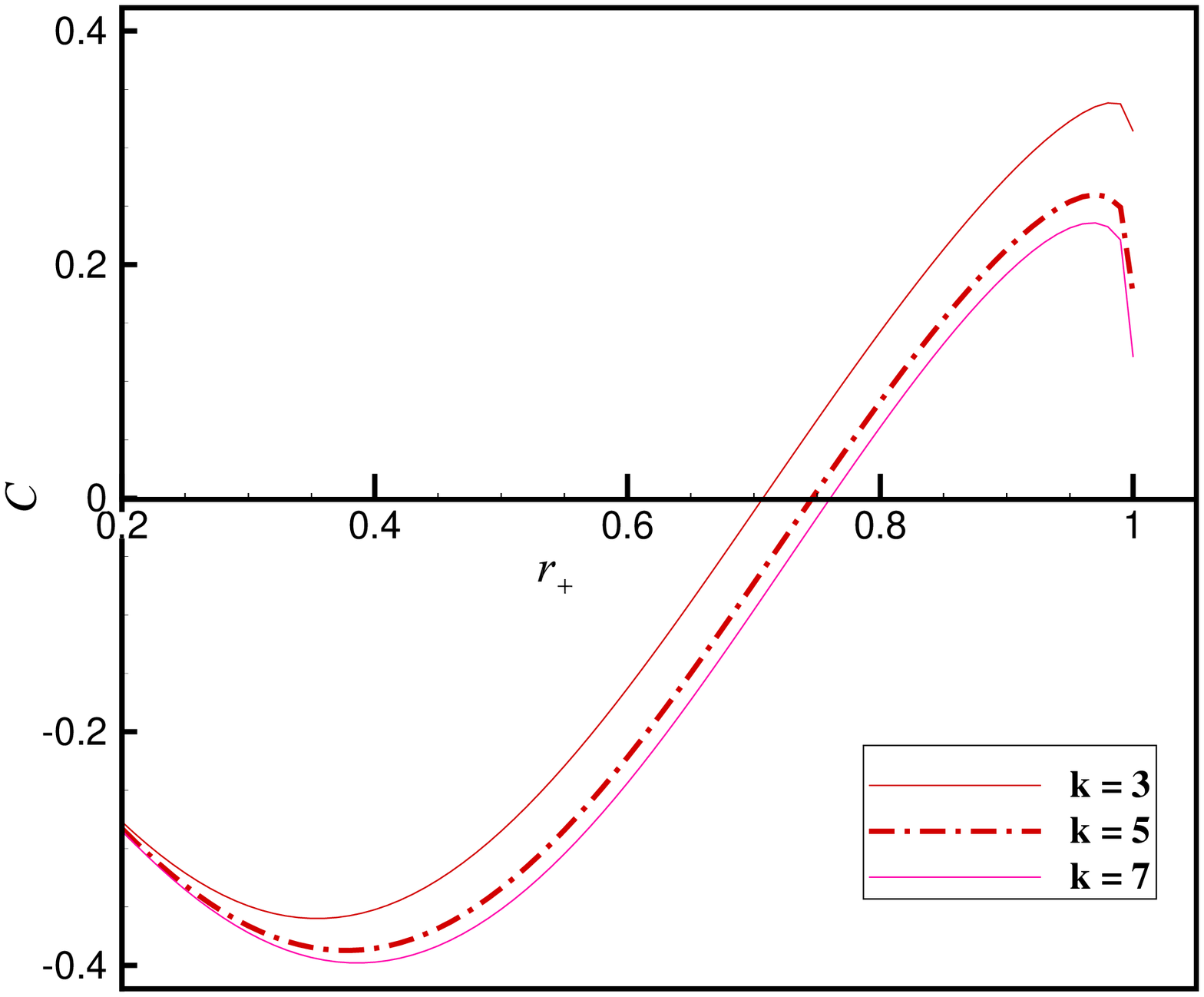}\label{fig4a}} \hspace*{.1cm}
\subfigure[\, $T$  versus $r_{+}$
]{\includegraphics[scale=0.25]{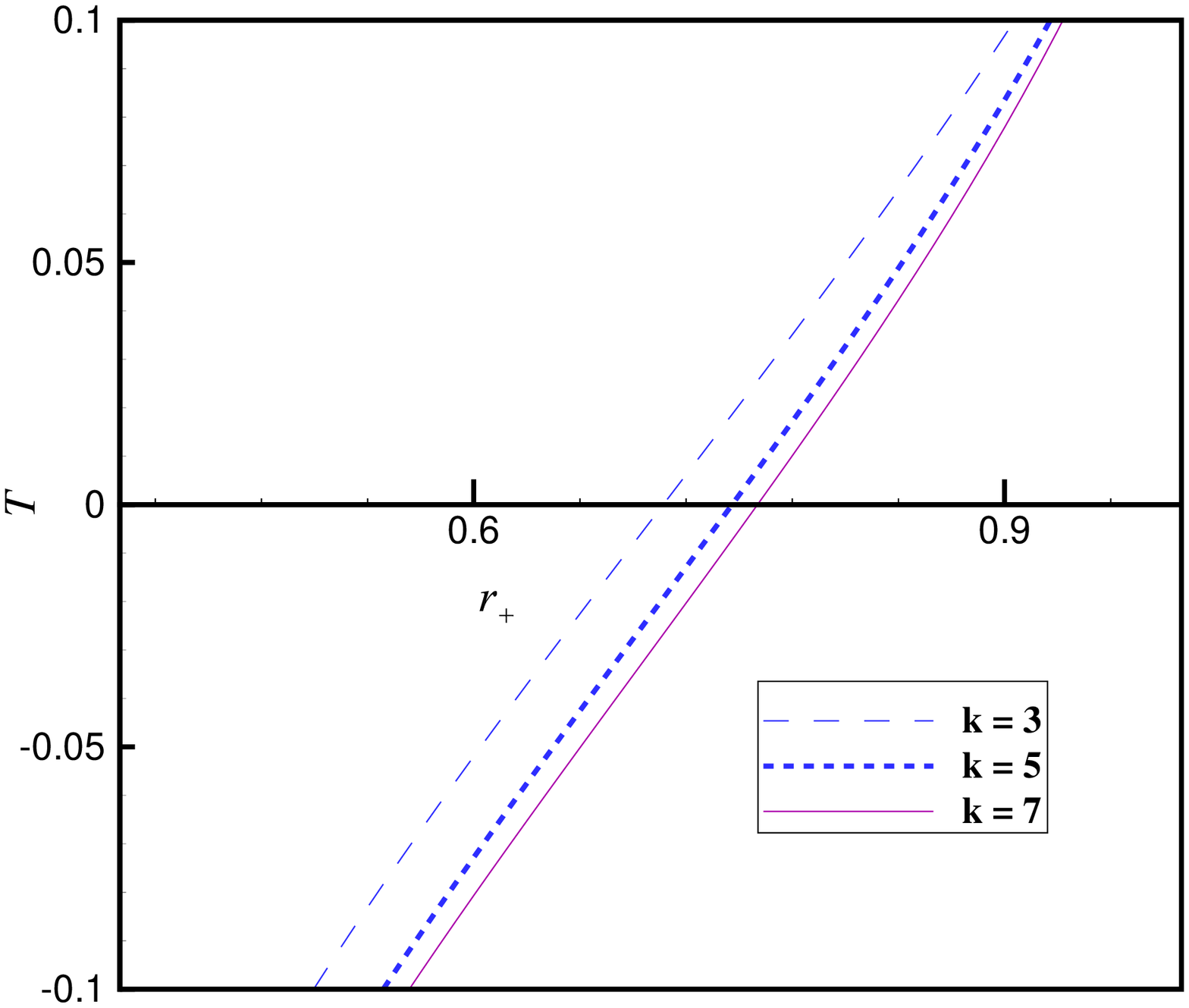}\label{fig4b}} \hspace*{.1cm}
\subfigure[\,$\rho_{+}$ versus $r_{+}$
]{\includegraphics[scale=0.25]{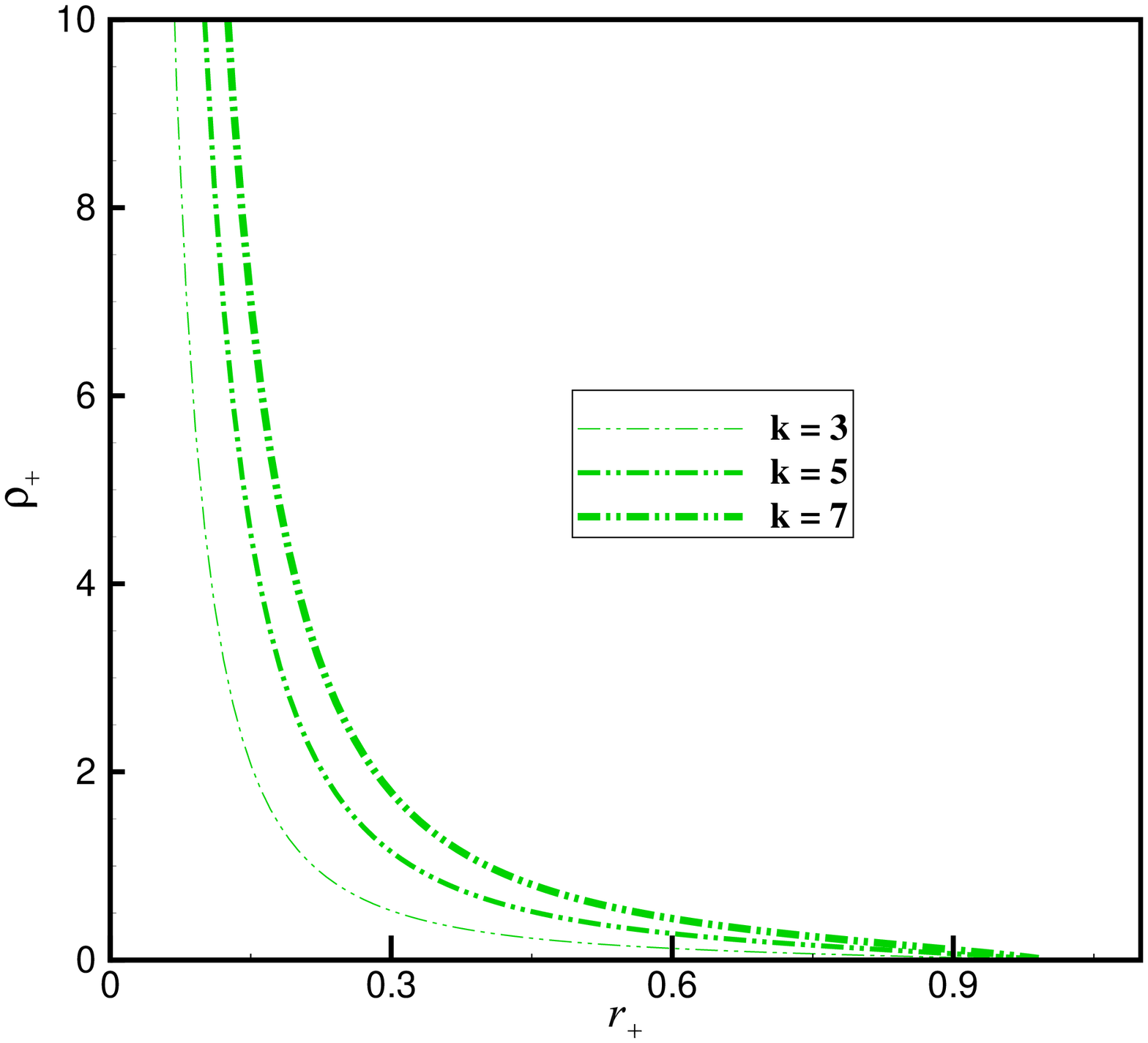}\label{fig4c}}
\caption{Behavior of $C$, $T$ and $\protect\rho_{+}$ respect to $r_{+}$ for $%
\mathbb{G}=l=1$ and $M=0.25$}
\label{fig4}
\end{figure}
\begin{table}[h]
\caption{Admissible domain (AD) for $\mathbb{G}=1$: $l=1$, $k=5${\protect\small {%
\ (left table)}, $M=0.17$, $k=5${\ (middle table) } and $M=0.5$, $l=1$ {%
(right table)}}}
\label{table}\centering
\begin{tabular}{|c|c|c|c|c|}
\hline\hline \toprule $M$ & $r_{\min}$ & $r_{\max}$ & AD &
$r_{_{{\tiny {div}}}}$ \\ \hline
0.17 & 0.48 & 0.60 & 0.12 & 0.29+0.56 I \\
0.18 & 0.52 & 0.66 & 0.14 & 0.34+0.59 I \\
0.19 & 0.56 & 0.72 & 0.16 & 0.37+0.60 I \\
0.20 & 0.60 & 0.77 & 0.17 & 0.40+0.60 I \\
0.21 & 0.63 & 0.82 & 0.19 & 0.42+0.63 I \\
0.22 & 0.67 & 0.87 & 0.20 & 0.44+0.65 I \\
0.23 & 0.69 & 0.92 & 0.23 & 0.46+0.66 I \\
0.24 & 0.72 & 0.96 & 0.24 & 0.48+0.68 I \\
0.25 & 0.74 & 1.00 & 0.26 & 0.49+0.69 I \\
0.26 & 0.77 & 1.04 & 0.27 & 0.51+0.70 I \\ \hline
\end{tabular}
\qquad 
\begin{tabular}{|c|c|c|c|c|}
\hline\hline \toprule $l$ & $r_{\min}$ & $r_{\max}$ & AD &
$r_{_{{\tiny {div}}}}$ \\ \hline
1.0 & 0.48 & 0.6 & 0.12 & 0.31+0.57 I \\
1.2 & 0.58 & 0.72 & 0.14 & 0.37+0.69 I \\
1.4 & 0.67 & 0.84 & 0.17 & 0.43+0.80 I \\
1.6 & 0.77 & 0.96 & 0.19 & 0.49+0.91 I \\
1.8 & 0.86 & 1.08 & 0.22 & 0.56+1.03 I \\
2.0 & 0.96 & 1.20 & 0.24 & 0.62+1.14 I \\
2.2 & 1.06 & 1.32 & 0.26 & 0.68+1.26 I \\
2.4 & 1.16 & 1.44 & 0.28 & 0.74+1.37 I \\
2.6 & 1.25 & 1.56 & 0.31 & 0.80+1.48 I \\
2.8 & 1.34 & 1.68 & 0.34 & 0.86+1.60 I \\ \hline
\end{tabular}
\qquad 
\begin{tabular}{|c|c|c|c|c|}
\hline\hline \toprule $k$ & $r_{\min}$ & $r_{\max}$ & AD &
$r_{_{{\tiny {div}}}}$ \\ \hline
3 & 1.09 & 1.73 & 0.64 & 0.66+0.84 I \\
4 & 1.12 & 1.73 & 0.61 & 0.70+0.85 I \\
5 & 1.14 & 1.73 & 0.59 & 0.72+0.86 I \\
6 & 1.16 & 1.73 & 0.57 & 0.49+0.91 I \\
7 & 1.17 & 1.73 & 0.56 & 0.74+0.87 I \\
8 & 1.17 & 1.73 & 0.56 & 0.75+0.87 I \\
9 & 1.18 & 1.73 & 0.55 & 0.75+0.87 I \\
10 & 1.18 & 1.73 & 0.55 & 0.76+0.87 I \\
11 & 1.18 & 1.73 & 0.55 & 0.76+0.87 I \\
12 & 1.19 & 1.73 & 0.54 & 0.76+0.87 I \\ \hline
\end{tabular}%
\end{table}
The existence of the stability range for black holes in this model
does not occur only in case $k=2$. To better investigate this
claim, the behavior of the heat capacity, temperature and
$\rho_{+}$ for different values of the parameter $k$ are depicted
in Fig. \ref{fig4}. According to these figures, we find that for
each value of parameter $k$, there is an admissible domain for the
black hole which decreases with increasing the value of $k$. To
clarify, one can see table \ref{table} (right panel). According to
this table, by increasing the value of $k$, the amount of
${r}_{\min }$ also increases while the amount of ${{r}_{\max }}$
is independent of the parameter $k$ and hence, the admissible
domain gets smaller. It should be noted that due to the imaginary
values of the heat capacity and the temperature as well as the
energy density at the horizon, the corresponding plots cannot be
continued much further after ${r}_{\max }$ and that's why they are
interrupted.

\begin{figure}[!htb]
\includegraphics[scale=0.4]{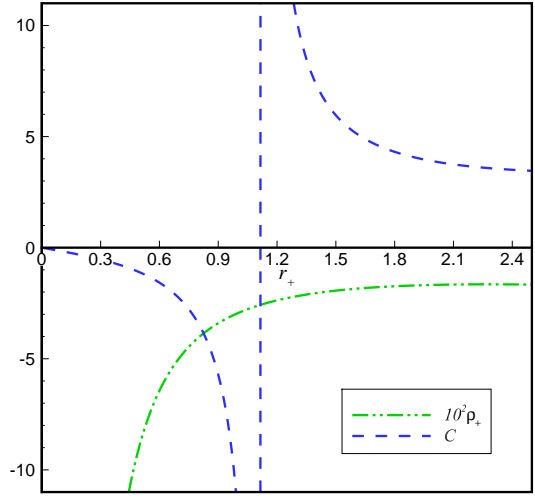}
\caption{Behavior of $C$ and $\protect\rho_{+}$ respect to $r_{+}$ for $k=2$%
, $l=\mathbb{G}=1$ and $M=1/12$}
\label{fig7}
\end{figure}

It is worthwhile to investigate the possibility of phase transition of the
proposed solution by studying the divergencies of the heat capacity. Our
calculations show that for the allowed region of the mass parameter, i.e. $M>%
\frac{1}{8\,\mathbb{G}}$, the heat capacity does not diverge in
the admissible domain of the black hole. This claim is supported
by the numerical results mentioned in the fourth column of the
table \ref{table}. To find these results, we have used Eq.
\eqref{heat} to calculate the radius in which function of $C$
diverges $(r_{_{{\tiny {div}}}})$ for different values of free
parameters $M$, $k$ and $l$. As can be seen from these tables, all
the radii $(r_{_{{\tiny {div}}}})$ are imaginary which means that
the heat capacity never diverges for the black holes whose radius
of the event horizon is in the admissible domain. It should be
mentioned that the divergencies of the heat capacity for the real
values of $r_{+}$ will occur if the mass parameter is out of the
allowed region. For more clarifications, we have plotted the
function $C$ together with the energy density at the horizon
versus $r_{+}$ in Fig. \ref{fig7}. Based on this figure, for the
real value of radius in which the heat capacity diverges, the
energy density is negative and hence, for the allowed region of
the mass parameter and the energy density, the solution is stable
and divergencies do not appear.


\subsubsection{Global thermal stability in grand canonical ensemble }

The idea of studying the black hole global stability was first
suggested by  Hawking and Page \cite{Page}. According to their
suggestion, the global stability of the black holes can be
explored with the help of a useful thermodynamic quantity, i.e.
Gibbs free energy.

Generally, the behavior of the Gibbs free energy particularly as a
function of thermodynamic parameters such as temperature, which
was first classified by Paul Ehrenfest, can indicate a phase
transition and some of the thermodynamic properties of different
phases \cite{phase}. Under Ehrenfest's scheme, phase transition
occurs when Gibbs free energy or at least one of its derivatives
with respect to one of its variables is discontinuous. More
clearly, phase transitions are labeled by the lowest derivative of
the Gibbs free energy which is discontinuous at the transition.
The existence of discontinuity in the first derivative of the
Gibbs free energy with respect to some thermodynamic parameters
indicates first-order phase transitions \cite{first}. In the case
of second-order phase transitions, Gibbs function and its first
derivative are continuous while the second derivative of the Gibbs
free energy meets discontinuity \cite{first}. According to the
Ehrenfest classification, third, fourth, and higher-order phase
transitions could in principle occur.

The black hole Gibbs free energy in the grand canonical ensemble,
in terms of the mass, temperature and entropy  is given by
\begin{equation}
    G=M-TS.
\end{equation}
It is well-known that the black hole with positive temperature is
globally stable provided that its Gibbs free energy is positive.

Regarding the above-mentioned point, one can use Eqs.
\eqref{entropy} and \eqref{tem} to obtain the functional form of
the Gibbs free energy with the following relation%
\begin{equation}
G\,=\,M+2\,\mathbb{G}\,M\left( k-1\right) \left[ 1-\left( 1-\eta \right) ^{{%
\frac{k}{k-1}}}\right] +\frac{1}{4}\left[ 1-k\left( 1+\frac{r_{+}^{2}}{l^{2}}%
\right) \right].
\end{equation}

\begin{figure}
    \centering\fbox{%
        \subfigure[\, $G$ versus
        $r_{+}$]{\includegraphics[scale=0.30]{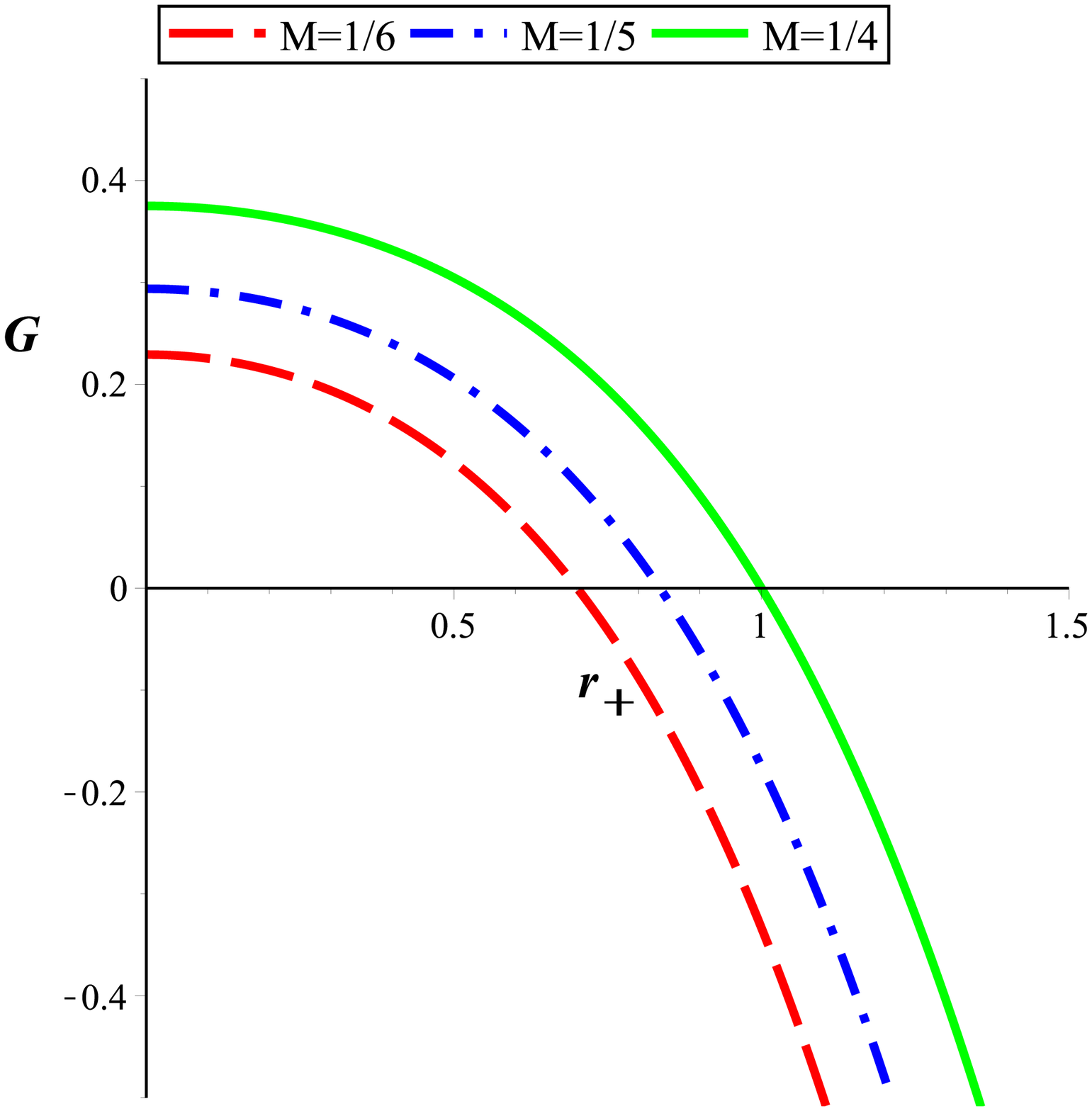}\label{fig5-1}}} \hspace*{.1cm} \fbox{%
        \subfigure[\, $G$  versus $T$
        ]{\includegraphics[scale=0.29]{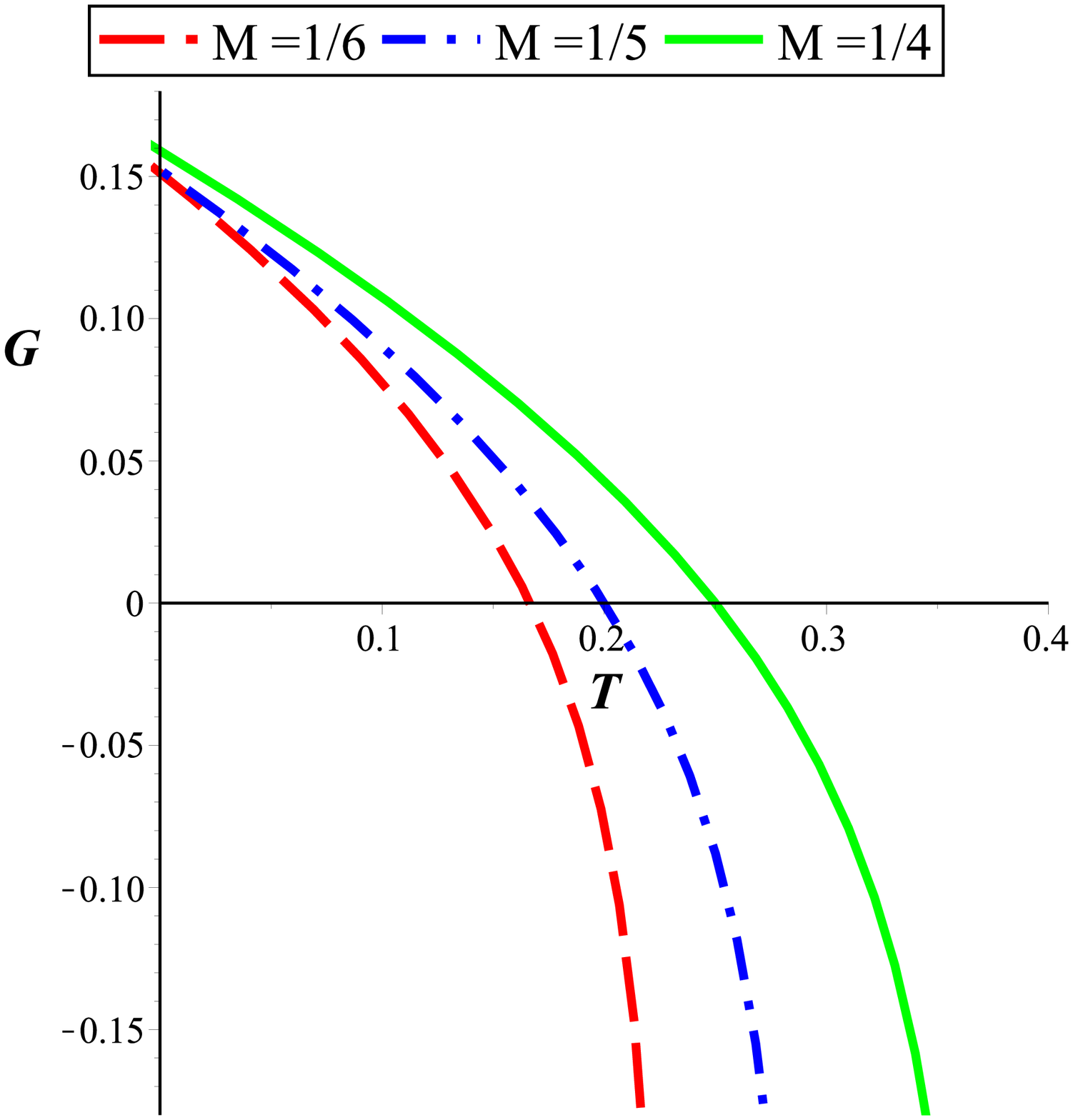}\label{fig5-2}}}
    \caption{Behavior of Gibbs free energy versus $r_{+}$ (left) and $T$ (right) for
        $\mathbb{G}=l=1$ and $k=2$}
    \label{fig5}
\end{figure}

Due to the complexity of the above function, it is not possible to
study its behavior analytically and, therefore, we use the
numerical solution to analyze its behavior. Examination of a wide
range of parameters shows that black holes with the event horizon
in the interval ${r}_{\min }<r_{+}<{r}_{\max } $ are globally
stable owing to the strictly decreasing behavior of the Gibbs free
energy functions. For instance, the behavior of Gibbs free energy
in terms of $r_{+}$ and $T$ for $k=2$ are sketched in Fig.
\ref{fig5} for different values of the mass parameter.\par To
ensure the above result, we calculate the first and second
derivative of the Gibbs free energy which will be as follows
\begin{equation}
G^\prime=\frac{dG}{dr_{+}}=\frac {kr}{{2\,l}^{2}} \left[  \left( 1-\frac{1}{8\,\mathbb{G}\,M}{ {\left[1+\frac{r_{+}^2}{l^2}\right]}} \right) ^{\frac{1}{ k-1 }}-1 \right] ,
\end{equation}
\begin{equation}
G^{\prime\prime}=\frac{d^2G}{dr_+^2}=\frac {k}{2{l}^{2}} \Bigg{\{}\left( 1-{\frac {1}{
        8\,\mathbb{G}\,M} \left[ 1+{\frac {{r_{+}}^{2}}{{l}^{2}}\frac{k+1 }{ k-1 }} \right] } \right) \left( 1-{\frac{1}{8\,\mathbb{G}\,M}{ {\left[1+\frac{r_{+}^2}{l^2}\right]}} } \right) ^{-{\frac {k-2}{k-1}}} -1 \Bigg{\}}.
\end{equation}
\begin{figure}
    \centering\includegraphics[scale=0.4]{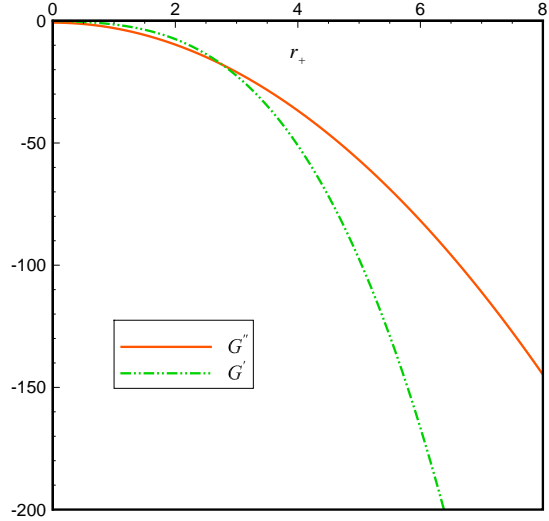}\caption{Behavior of $G^\prime$ and $G^{\prime\prime}$ versus $r_{+}$ for $\mathbb{G}=1$, $M=1/5$, $l=3$ and $k=2$}\label{dG}
\end{figure}
We try to analyze the behavior of the above functions using the
numerical solution. The choice of different sets of parameters
indicate that the first and second derivatives of Gibbs free
energy are smooth functions with respect to $r_+$ which as an
example, the result of a set of parameters is shown in
Fig.\ref{dG}. In the same way, one can show that the higher
derivatives of $G$ with respect to $r_+$ also behave similarly and
they are smooth functions. The smoothness of the Gibbs free energy
along with its derivatives guarantee the global stability of the
black hole.
 \begin{figure}[tbh]
    \centering\includegraphics[scale=0.5]{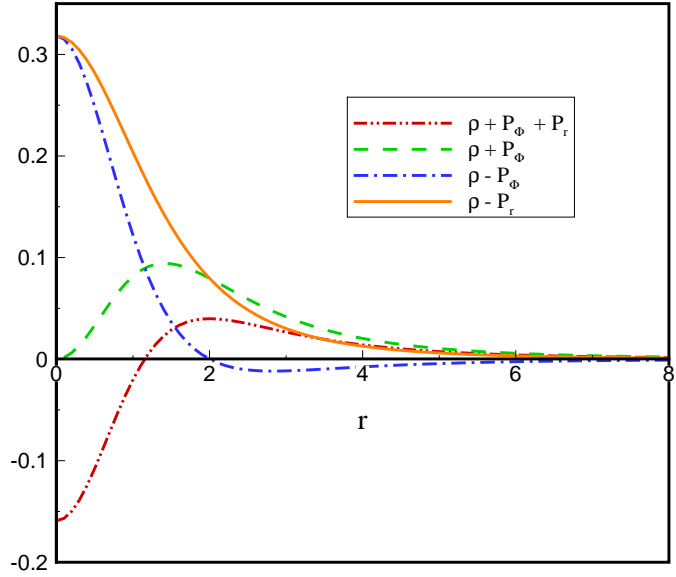}\caption{Energy conditions for $\mathbb{G}=l=L=1$, $M=2$ and $k=2$}\label{energy}
\end{figure}

Before ending this section, it is worth studying the energy
condition of the proposed black hole. To this end, we recall that
the singularity theorems of Hawking and Penrose establish the
relation between the appearance of singularities inside the black
holes and the validity of the SEC \cite{Penrose}. However,
avoidance of  Hawking and Penrose singularity theorem to construct
regular black holes was first explained by Borde in 1997
\cite{Borde:1996df}. Following Borde's theorem, Mars, Martin-Prats
and Senovilla considered the spherically symmetric and static
spacetime and proved that if these spacetimes are regular at the
origin and satisfy the SEC, they cannot include any black hole
region in GR \cite{Mars}. Therefore, reversing the singularity
theorem of Hawking and Penrose leads to the conclusion that
{\it{"regular black holes violate the SEC somewhere inside the
horizon"}} \cite{Bargueno:2020ais}. It is interesting to note that
the violation of the SEC inside the event horizon received a
simple formulation in terms of the Tolman mass which is considered
as a clear criterion to evaluate the degree of such violation
\cite{Zaslavskii:2010qz}. \par To study the energy condition for
our suggested model in the obtained admissible domain, we make use
of Eq. (\ref{emtensor}) and we receive the following results
\cite{energy condition}
\begin{eqnarray}
&&SEC=\rho+P_r+P_{\Phi}\geqslant 0\,,\label{SEC}\\
&&NEC_{1,2} \equiv WEC_{1,2}=\rho+P_{r,\Phi}\geqslant 0\,,\label{WEC}\\
&&DEC_3 \equiv WEC_3=\rho\geqslant 0\,,\label{WE3}\\
&&DEC_{1,2}=\rho-P_{r,\Phi}\geqslant 0\,.\label{DEC}
\end{eqnarray}
According to Eqs. \eqref{rho} and \eqref{Pt}, one can easily find
that just for areas with radii larger than
$\sqrt{\frac{2LM}{2k-1}}$, the SEC is satisfied that actually
corresponds to the region in which the tangential pressure is
positive. Moreover, due to the positive values of the mass
parameter and $L$, the WEC is met everywhere. This is while,
regarding  the DEC, our calculations show that it is violated in
some areas. In fact, for regions with ${r}_{\min
}<r_{+}<\sqrt{\frac{2LM}{k-1}}$, DEC is satisfied while for the
rest of the admissible domain,
$\sqrt{\frac{2LM}{k-1}}<r_{+}<{r}_{\max }$, DEC will be violated.
To be more clear, we have provided Fig.\ref{energy} in which the
behavior of the combination of the energy-momentum tensor
components (various energy conditions) in terms of $r$ are
depicted.

\section{Concluding remarks}\label{conclusion}

In this paper, we proposed a new model of regular black hole based
on considering a new model of energy density in {$d-$dimensions}
that follows requirements mentioned in reference
\cite{Aros:2019quj}. This suggested function is strictly
decreasing, having the maximum value at the origin, which leads to
avoiding the central singularity, and zero at infinity. {In this
paper we focused on the $(2+1)-$ dimensional solutions and
studying the properties of the solutions in the higher dimensions
was left for future works.} Studying the behavior of the proposed
energy density on the event horizon showed that there is an upper
limit on the radius of the event horizon of such black holes which
is completely compatible with the condition of the black holes
whose mass is {finite}. Moreover, we understood that to have a
real value for the radius of the event horizon,
the mass parameter of the black hole must be larger than $\frac{1}{8\,%
\mathbb{G}}$.

Regarding the obtained solution, we found that, depending on the
metric parameters, this solution could represent a black hole with
two horizons or an extreme black hole. By selecting a certain
value of $L$ one could get a flat, dS or AdS core. Also, the
thermodynamics of the proposed solution was studied and the first
law of black hole thermodynamics was checked.

Next, we studied the thermal stability in the canonical ensemble.
In this regard, in addition to the positivity of the temperature
and heat capacity, our proposed model also required an additional
condition that is related to the positivity of the energy density.
To be more accurate, the behavior of temperature, heat capacity
and energy density were studied for different values of the model
parameters. We concluded that for the stable black holes the
radius of the event horizon {should be selected} in the interval
$r_{\min}<r_{+}<r_{\max}$. The critical behavior of the obtained
solution was investigated by studying the divergencies of the heat
capacity. We
found that the radius in which the function of $C$ diverges $(r_{_{{\tiny {%
div}}}})$ is not in the admissible domain of the black hole. Also,
we investigated the global stability of the solution by studying
the Gibbs free energy. Our investigation showed that the black
holes with the radius of the event horizon in the mentioned
interval are globally stable and hence, we {conjecture} they do
not experience any kind of classical thermal phase transitions. In
the end, the energy conditions were checked for this solution. We
concluded that although the WEC is met everywhere, the SEC, as
well as the DEC, will be violated for a part of the admissible
domain of the event horizon radius.

It is interesting to investigate the dynamic stability of the
solutions and analyze quasi-normal modes. Besides, it is nice to
study the causal structure of the obtained solution via the
possible Penrose diagrams. Moreover, one can examine the effect of
perturbations on the Cauchy horizon stability. Also, more
geometrical/topological investigations and looking for topological
defects can be regarded. All these interesting suggestions can be
addressed in independent work.


\section*{Acknowledgments}

We thank Shiraz University Research Council. S. Mahmoudi is
grateful to the Iran Science Elites Federation for the financial
support.

\appendix \section{ A review on some regular black hole models}\label{appendix1}

In this appendix, we try to review some efforts that have been
done to propose curvature singularity-free models of black holes.

Curvature singularities of GR could be understood as points where
every classical theory of gravity does not make sense. However,
the important point is that classical theories of gravity cannot
be valid at all scales \cite{DeLorenzo:2014pta} and, therefore,
describing nature at some scales such as Planck scales requires a
new theory of gravity. In the context of these new theories, such
as string theory or Loop Quantum Gravity (LQG), the problem of
singularity can be solved by considering quantum effects. For
instance, the results of LQG indicate that quantum gravity
fluctuations produce enough pressure to counteract the
gravitational effect before the matter reaches the Planck density.
In connection with black holes, this scenario suggests that the
gravitational collapse terminates before a singularity is formed.
This process leads to the formation of a dense central core whose
density is of the order of the Planck density. These objects are
called Planck stars which exist within a black hole's event
horizon \cite{Rovelli:2014cta}. It is necessary to mention that
since the starting point for the emergence of the
quantum-gravitational effects is controlled by energy density and
not by size, a Planck star is calculated to be much larger than
the Planck scale \cite{Rovelli:2014cta}.

Black holes whose inside contain a dense core instead of a
singularity could be regarded as regular (or non-singular) black
holes. In practice, regular black holes can be studied as a
geometry that recovers a standard black hole solution at distances
sufficiently far from the core while whose center can be treated
as a manifold. The general idea for providing a suitable model to
describe these types of black holes is changing the mass parameter
into a radial mass function such that near the origin the mass
function behaves in a way that singularity disappears and the
solution would be regular. More clearly, the regularity of the
energy density at the center in $d$ dimensional spacetime requires
the mass function $m(r)$ to vanish as $r^{d-1}$ when $r$ tends to
zero. In this way, the first idea was proposed by  Sakharov and
Gline which based on their suggestion singularities could be
avoided by a non-singular dS core, with the equation of state
$p=-\rho$ \cite{sakharov, Gliner}. Following this idea, the first
regular black holes solution was proposed by Bardeen in which
there are horizons but no singularity and close to the origin,
solution meets a dS geometry \cite{Bardeen}. The Bardeen model is
described by the following metric
    \begin{equation}
    \text{$ds^2$}=-\left( 1-\frac{2mr^2}{(r^2+L^2)^{3/2}}\right) \text{%
        $dt$}^2+\left( 1-\frac{2mr^2}{(r^2+L^2)^{3/2}}\right) ^{-1}\text{%
        $dr$}^2+r^2\text{$d\Omega $}^2,  \label{Bardeen}
    \end{equation}
where $L$ has the role of a regulator to avoid the presence of a
singularity. Calculation of the curvature invariants shows that
this model characterizes the regular spacetime. To investigate the
physical interpretation for parameters $m$ and $L$, we can examine
the asymptotic behavior of the metric which will be as follows
    \begin{equation}
    g_{tt}=-1+2m/r-3mL^2/r^3+O(1/r^5).
    \end{equation}
Since the second term goes as $1/r$ the parameter $m$ will be
associated with the mass of the configuration. However, the next
term changes as $1/r^3$ and thus we are not allowed to relate the
parameter $L$ with some kind of charge like, for instance, in the
Reissner--Nordstr\"om solution. A physical source associated with
Bardeen's solution was clarified nearly thirty years later,  when
Ayon-Beato and Garcia \cite{Ayon-Beato:2000mjt} successfully
interpreted Bardeen's black hole in the context of nonlinear
electrodynamics and found that $L$ can be interpreted as the
monopole charge of a self-gravitating magnetic field described by
nonlinear electrodynamics.

As it was mentioned before, considering the quantum gravity
corrections can lead to the removal of the curvature singularities
existing in the standard black hole geometries. However, since the
theory of quantum gravity is not available, regular black holes
issues can be regarded as phenomenological toy models in order to
explore possible ways to solve the problem of singularity. In
fact, quantum gravity corrections can be imitated by introducing
an anisotropic fluid that must satisfy a set of conditions. It
means that introducing an anisotropic fluid that strongly
concentrates at the origin could have the same results to
eliminate the singularity. The logic behind this statement is that
geometry whose source is the mentioned anisotropic fluid could
effectively arise from a low energy limit of quantum gravity, as a
solution to Einstein's equations modified by quantum theory.
Therefore,  it is feasible to construct non-singular black hole
solutions including matter fields in the energy-momentum tensor.
In this regard, to provide an appropriate model of energy density
which leads to a regular black hole solution, the general
conditions must be satisfied by the energy density which will be
discussed in what follows.

Here, for simplicity, we impose a highly symmetric geometry and
consider only the static case. A $d-$dimensional static spherical
symmetric geometry in Schwarzschild coordinates can be described
by
    \begin{equation}\label{ Sch.coordinate}
    ds^2 =-f(r) dt^2+ \frac{dr^2}{f(r)} + r^2 d \Omega^2_{d-2}.
    \end{equation}

Moreover, an anisotropic fluid living in this spacetime,
considering the symmetries of the geometry, must have the
following form
    \begin{equation}\label{anisotropic fluid}
    T^{\mu}_{\hspace{1ex}\nu} =  \textrm{diag}(-\rho, p_r, p_\theta, p_\theta, ...),
    \end{equation}
where  $\rho=-p_r$, due to the consistency with the Einstein field
equations. It is notable that the negative value of the radial
component of the pressure can explain the regularity of the
solution in the sense that it can impede collapse by weakening the
gravitational field \cite{Brustein:2018web}. Besides, the
conservation law $\nabla_{\mu} T^{\mu\nu} =0$ implies that
    \begin{equation}
    p_\theta = \frac{r}{d-2} \frac{d}{dr}p_r + p_r. \label{pressure}
    \end{equation}
Now, to examine the general behavior of the energy density, it is
convenient to define the mass function as follows
\cite{Aros:2019quj}
    \begin{equation}\label{energy density}
    m(r)=\Omega_{d-2} \int_0^r  \rho(x) x^{d-2} dx.
    \end{equation}
In order to have a well-defined physical solution, the suggested
energy density must meet the requirements described in what
follows \cite{Aros:2019quj}:
    \begin{enumerate}
        \item{Although singularities could be replaced by regular regions including the matter that might violate the SEC, the WEC must be satisfied and therefore the energy density must be positive.
        }
        \item{$\rho$ must be a continuously differentiable function to avoid singularities. This point requires that mass function ($m(r)$), is a positive monotonically increasing function, i.e. if $r_1\geq r_2$ then $m(r_1)\geq m(r_2)$,
            which vanishes at the origin.}
        \item{To guarantee a well posed asymptotic behavior, the energy density should be a decreasing function of radial coordinate to vanish at spatial infinity. In fact, $\rho(r)$ must be such that $m(r)$ tends to a constant M which is proportional to the mass of the solution, i.e.
            \begin{eqnarray}
            \lim_{r \rightarrow \infty} \rho(r) =0,\nonumber\\
            \lim_{r \rightarrow \infty} m(r) = {M},
            \end{eqnarray}
            which means that
            \begin{equation}
            \lim_{r \rightarrow \infty} \frac{d}{dr} m(r) = 0.
            \end{equation}}
        \item{In order to intimate the quantum gravitational effects, the energy density $\rho$ must have a single maximum at the origin ($r = 0$) and rapidly decreases away from the center which yields the following condition
            \begin{equation}
            \left. m(r) \right|_{r\approx 0}  \approx K r^{d-1},
            \end{equation}
            where $K$ is a positive constant and proportional to $\rho(0)$. The finiteness of $\rho(0)$ guarantees absence of singularity at $ r = 0$. Moreover, $\rho$ must be such that there is a radious $r=R$ in which $m(R) \approx M$ and $\frac{d}{dr} m(r)|_{r=R} \approx 0$. Generally speaking, $R$ could be in the interval $\ell_P \ll R \ll r_{+}$ for large masses where  $\ell_P $ and $r_+$ stand for Planck length and horizon radius, respectively. However, this condition cannot be applied to configuration whose mass is within the range of Planck scale \cite{I. Dymnikova}.
        }
    \end{enumerate}

Regarding the mentioned criteria, Estrada and Aros proposed a
$d-$dimensional model of energy density to describe a non-singular
black hole as follows \cite{Aros:2019quj}
    \begin{equation}
    \rho(r) = \frac{d-1}{\Omega_{d-2}} \frac{L^{d-2}M^2}{(L^{d-2}M+r^{d-1})^2},
    \end{equation}
where in $4-$dimensions, it reduces to the Hayward metric, a
minimal model of energy density to describe Planck stars
\cite{Hayward:2005gi}. It should be noted that, here, $L$ plays
the role of the regulator to prevent the formation of singularity.

However, the above relation for the energy density is not the most
general form that meets all the mentioned criteria. Therefore,
motivated by the model suggested by Estrada and Aros, we propose
the following energy density
    \begin{equation}
    \rho(r) =\, \frac{d-1}{\Omega_{d-2}}\,\frac{k-1}{2\,L^{d-2}} (1+\frac{r^{d-1}}{2\,L^{d-2}M})^{-k},
    \end{equation}
where $k\geq 2$ is an integer number. Our suggestion could be
viewed as a generalization of the model was put forward by Estrada
and Aros which for $k=2$ and $d=4$ coincides with the Hayward
model.

Finally, it should be noted that although the above energy density
is inspired by a quantum gravity model, it can effectively be used
as a classical model to describe regular black holes.



\begin{thebibliography}{99}
\bibitem{LIGOScientific:2016aoc} B.~P.~Abbott \textit{et al.} [LIGO
Scientific and Virgo],
Phys. Rev. Lett. \textbf{116}, 061102 (2016)

\bibitem{LIGOScientific:2016sjg} B.~P.~Abbott \textit{et al.} [LIGO
Scientific and Virgo],
Phys. Rev. Lett. \textbf{116}, 241103 (2016)

\bibitem{LIGOScientific:2017bnn} B.~P.~Abbott \textit{et al.} [LIGO
Scientific and VIRGO],
Phys. Rev. Lett. \textbf{118}, 221101 (2017) [erratum: Phys. Rev.
Lett. \textbf{121}, 129901 (2018)]

\bibitem{EventHorizonTelescope:2019dse} K.~Akiyama \textit{et al.} [Event
Horizon Telescope],
Astrophys. J. Lett. \textbf{875}, L1 (2019)

\bibitem{penrose} R. Penrose,\textit{"Singularities of spacetime," in
Theoretical Principles in Astrophysics and Relativity}, N. R. Lebovitz, W.
H. Reid, and P. O. Vandervoort, Eds., vol. 1, pp. 217-243, University of
Chicago Press, Chicago, Ill, USA, 1978.

\bibitem{Joshi:2013xoa} P.~S.~Joshi, 
[arXiv:1311.0449]

\bibitem{PH1} R. Penrose, Phys. Rev. Lett. \textbf{14}, 57 (1965)

\bibitem{PH2} S. W. Hawking, Proc. Roy. Soc. A \textbf{295}, 490 (1966)

\bibitem{PH3} S. W. Hawking, Proc. Roy. Soc. A \textbf{294}, 511 (1966)

\bibitem{PH4} S. W. Hawking, Proc. Roy. Soc. A \textbf{300}, 187 (1967)

\bibitem{PH5} S. W. Hawking and R. Penrose, P. Roy. Soc. A \textbf{314}, 529
(1970)  

\bibitem{NSBHs1} H. A. Buchdahl, Mont. Not. Roy. Astron. Soc. \textbf{150},
1 (1970)  

\bibitem{NSBHs2} A. A. Starobinsky, Phys. Lett. B \textbf{91}, 99 (1980)

\bibitem{NSBHs3} A. Borisov, B. Jain, and P. Zhang, Phys. Rev. D \textbf{85}%
, 063518 (2012)  

\bibitem{NSBHs4} G. J. Olmo and D. Rubiera-Garcia, Phys. Rev. D \textbf{84},
124059 (2011)  

\bibitem{NLEDNSBHs} C. Corda and H. J. M. Cuesta, Mod. Phys. Lett. A \textbf{%
25}, 2423 (2010)  

\bibitem{PlanckStars1} T. De Lorenzo, C. Pacilio, C. Rovelli and S.
Speziale, Gen. Rel. Grav. \textbf{47}, 41 (2015) 

\bibitem{PlanckStars2} C. Rovelli and F. Vidotto, Int. J. Mod. Phys. D
\textbf{23}, 1442026 (2014) 

\bibitem{sakharov} A. D. Sakharov, Zh. Eksp. Teor. Fiz. \textbf{49},
345 [Sov. Phys. JETP22, 241 (1966)]

\bibitem{Gliner} E. B. Gliner, Sov. Phys. JETP \textbf{22}, 378 (1966)

\bibitem{Bardeen} J. M. Bardeen,
in Proceedings of International Conference GR5 (Tbilisi, USSR,
1968) p. 174.
%

\bibitem{Ayon-Beato:1998hmi} E.~Ayon-Beato and A.~Garcia,
Phys. Rev. Lett. \textbf{80}, 5056 (1998)

\bibitem{Bronnikov:2000vy} K.~A.~Bronnikov,
Phys. Rev. D \textbf{63}, 044005 (2001)

\bibitem{Dymnikova:2004zc} I.~Dymnikova,
Class. Quant. Grav. \textbf{21}, 4417 (2004)

\bibitem{Novello:1999pg} M.~Novello, V.~A.~De Lorenci, J.~M.~Salim and
R.~Klippert,
Phys. Rev. D \textbf{61}, 045001 (2000)

\bibitem{Culetu:2014lca} H.~Culetu,
Int. J. Theor. Phys. \textbf{54}, 2855 (2015)

\bibitem{Balart:2014cga} L.~Balart and E.~C.~Vagenas,
Phys. Rev. D \textbf{90}, 124045 (2014)

\bibitem{Hayward:2005gi} S.~A.~Hayward,
Phys. Rev. Lett. \textbf{96}, 031103 (2006)

\bibitem{Azreg-Ainou:2014pra} M.~Azreg-A\"\i{}nou,
Phys. Rev. D \textbf{90}, 064041 (2014)

\bibitem{Sajadi} S. H. Hendi, S. N. Sajadi and M. Khademi,
 Phys. Rev. D \textbf{103}, 064016 (2021)

\bibitem{Zaslavskii:2010qz} O.~B.~Zaslavskii,
Phys. Lett. B \textbf{688}, 278 (2010)

\bibitem{Nicolini:2005vd} P.~Nicolini, A.~Smailagic and E.~Spallucci,
Phys. Lett. B \textbf{632}, 547 (2006)

\bibitem{Dymnikova:1992ux} I.~Dymnikova,
Gen. Rel. Grav. \textbf{24}, 235 (1992) 

\bibitem{Spallucci:2017aod} E.~Spallucci and A.~Smailagic,
Int. J. Mod. Phys. D \textbf{26}, 1730013 (2017)

\bibitem{Aros:2019auf} M.~Estrada and R.~Aros,
Eur. Phys. J. C \textbf{79}, 810 (2019)

\bibitem{Aros:2019quj} R.~Aros and M.~Estrada,
Eur. Phys. J. C \textbf{79}, 259 (2019)

\bibitem{Babichev:2020qpr} E.~Babichev, C.~Charmousis, A.~Cisterna and
M.~Hassaine,
JCAP \textbf{06}, 049 (2020)

\bibitem{Ayon-Beato:1999kuh} E.~Ayon-Beato and A.~Garcia,
Phys. Lett. B \textbf{464}, 25 (1999)

\bibitem{Ayon-Beato:1999qin} E.~Ayon-Beato and A.~Garcia,
Gen. Rel. Grav. \textbf{31}, 629 (1999)

\bibitem{Burinskii:2002pz} A.~Burinskii and S.~R.~Hildebrandt,
Phys. Rev. D \textbf{65}, 104017 (2002)

\bibitem{Ma:2015gpa} M.~S.~Ma,
Annals Phys. \textbf{362}, 529 (2015)
\bibitem{Carlip:1995qv} S.~Carlip, 
Class. Quant. Grav. \textbf{12}, 2853 (1995)
\bibitem{Welling:1995er}
M.~Welling,
Class. Quant. Grav. \textbf{13}, 653-680 (1996)
\bibitem{Banados:1992wn} M.~Banados, C.~Teitelboim and J.~Zanelli,
Phys. Rev. Lett. \textbf{69}, 1849 (1992)

\bibitem{Banados:1992gq} M.~Banados, M.~Henneaux, C.~Teitelboim and
J.~Zanelli, 
Phys. Rev. D \textbf{48}, 1506 (1993) [erratum: Phys. Rev. D \textbf{88}%
, 069902 (2013)]
\bibitem{Chan:1994qa} K.~C.~K.~Chan and R.~B.~Mann,
Phys. Rev. D \textbf{50}, 6385 (1994) [erratum: Phys. Rev. D \textbf{52},
2600 (1995)] 

\bibitem{Zaslavsky:1994dx} O.~B.~Zaslavsky,
Class. Quant. Grav. \textbf{11}, L33 (1994)



\bibitem{Martinez:1999qi} C.~Martinez, C.~Teitelboim and J.~Zanelli,
Phys. Rev. D \textbf{61}, 104013 (2000)

\bibitem{Yamazaki:2001ue} R.~Yamazaki and D.~Ida,
Phys. Rev. D \textbf{64}, 024009 (2001)

\bibitem{Gurtug:2010dr} O.~Gurtug, S.~H.~Mazharimousavi and M.~Halilsoy,
Phys. Rev. D \textbf{85}, 104004 (2012)

\bibitem{Hendi:2015bba} S.~H.~Hendi,
Gen. Rel. Grav. \textbf{48}, 50 (2016)
\bibitem{Tang:2019jkn} Z.~Y.~Tang, Y.~C.~Ong, B.~Wang and
E.~Papantonopoulos,
Phys. Rev. D \textbf{100}, 024003 (2019)
\bibitem{QGravity}
S.Carlip, 1998 Quantum Gravity in 2+1 Dimensions (Cambridge University Press, Cambridge, England)


\bibitem{Witten:1998zw}
E.~Witten,
Adv. Theor. Math. Phys. \textbf{2}, 505 (1998)
%
\bibitem{Witten:2007kt}
E.~Witten,
[arXiv:0706.3359]
\bibitem{Birmingham:2001pj}
D.~Birmingham, I.~Sachs and S.~N.~Solodukhin,
Phys. Rev. Lett. \textbf{88}, 151301 (2002)
\bibitem{Ren:2010ha}
J.~Ren,
JHEP \textbf{11}, 055 (2010)
\bibitem{Liu:2011fy}
Y.~Liu, Q.~Pan and B.~Wang,
Phys. Lett. B \textbf{702}, 94-99 (2011)
\bibitem{KordZangeneh:2017zyy}
M.~Kord Zangeneh, Y.~C.~Ong and B.~Wang,
Phys. Lett. B \textbf{771}, 235-241 (2017)
\bibitem{Hendi:2010px}
S.~H.~Hendi,
Eur. Phys. J. C \textbf{71}, 1551 (2011)



%
\bibitem{Anacleto:2014cga}
M.~A.~Anacleto, F.~A.~Brito and E.~Passos,
Phys. Lett. B \textbf{743}, 184 (2015)

\bibitem{Ghosh:2011tt}
S.~G.~Ghosh,
Int. J. Mod. Phys. D \textbf{21}, 1250022 (2012)
\bibitem{Hendi:2015uia}
S.~H.~Hendi, B.~Eslam Panah, M.~Momennia and S.~Panahiyan,
Eur. Phys. J. C \textbf{75}, 457 (2015)

%
\bibitem{Ayon-Beato:2009rgu}
E.~Ayon-Beato, A.~Garbarz, G.~Giribet and M.~Hassaine,
Phys. Rev. D \textbf{80}, 104029 (2009)
\bibitem{Hendi:2017mgb}
S.~H.~Hendi, B.~Eslam Panah, S.~Panahiyan and A.~Sheykhi,
Phys. Lett. B \textbf{767}, 214 (2017)
%
\bibitem{Hendi:2016pvx}
S.~H.~Hendi, B.~Eslam Panah and S.~Panahiyan,
JHEP \textbf{05}, 029 (2016)
%
\bibitem{Hendi:2016wwj}
S.~H.~Hendi, B.~Eslam Panah and S.~Panahiyan,
PTEP \textbf{2016}, 103A02 (2016)
%
\bibitem{Hendi:2016hbe}
S.~H.~Hendi, S.~Panahiyan, S.~Upadhyay and B.~Eslam Panah,
Phys. Rev. D \textbf{95}, 084036 (2017)




\bibitem{Balart:2014jia} L.~Balart and E.~C.~Vagenas,
Phys. Lett. B \textbf{730}, 14 (2014)

\bibitem{Fan:2016hvf} Z.~Y.~Fan and X.~Wang,
Phys. Rev. D \textbf{94},  124027 (2016)

\bibitem{Estrada:2020tbz} M.~Estrada and F.~Tello-Ortiz,
[arXiv:2012.05068]
%

\bibitem{Natarajan:2008ks} P.~Natarajan and E.~Treister,
Mon. Not. Roy. Astron. Soc. \textbf{393}, 838 (2009)
\bibitem{Aoki:2020prb}
S.~Aoki, T.~Onogi and S.~Yokoyama,
Int. J. Mod. Phys. A \textbf{36}, no.10, 2150098 (2021)

\bibitem{Bekenstein:1973ur} J.~D.~Bekenstein, 
Phys. Rev. D \textbf{7}, 2333 (1973) 

\bibitem{Hawking:1998jf} S.~W.~Hawking and C.~J.~Hunter,
Phys. Rev. D \textbf{59}, 044025 (1999)
\bibitem{Ma:2014qma}
M.~S.~Ma and R.~Zhao,
Class. Quant. Grav. \textbf{31}, 245014 (2014)
\bibitem{Estrada:2019qsu}
M.~Estrada and R.~Aros,
Eur. Phys. J. C \textbf{79}, no.10, 810 (2019)
\bibitem{Estrada:2019cig}
M.~Estrada and R.~Aros,
Eur. Phys. J. C \textbf{80}, no.5, 395 (2020)
\bibitem{Page}
Hawking, S.W., Page, D.N.
 Commun.Math. Phys. \textbf{87}, 577–588 (1983).
\bibitem{phase}
Jaeger, G.
Arch Hist Exact Sc. \textbf{53}, 51–81 (1998).
\bibitem{first}
Blundell, Stephen J.; Katherine M. Blundell (2008). Concepts in Thermal Physics. Oxford University Press. ISBN 978-0-19-856770-7.








\bibitem{Penrose}
S. W. Hawking and R. Penrose, Proc. Roy. Soc. Lond. A
{\bf{314}} , 529 (1970)
\bibitem{Borde:1996df}
A.~Borde,
Phys. Rev. D \textbf{55}, 7615-7617 (1997)
\bibitem{Mars}
M. Mars, M. M. Mart´n ı-Prats and J. M. M. Senovilla, Phys. Lett. A \textbf{218} , 147 (1996)
\bibitem{Bargueno:2020ais}
P.~Bargue\~no,
Phys. Rev. D \textbf{102}, no.10, 104028 (2020)













\bibitem{energy condition}
M. Visser, {\it Lorentzian wormholes: from Einstein to Hawking},
United Book Press, Springer-Verlag, New York (1995)
%
\bibitem{DeLorenzo:2014pta}
T.~De Lorenzo, C.~Pacilio, C.~Rovelli and S.~Speziale,
Gen. Rel. Grav. \textbf{47}, no.4, 41 (2015)
\bibitem{Rovelli:2014cta}
C.~Rovelli and F.~Vidotto,
Int. J. Mod. Phys. D \textbf{23}, no.12, 1442026 (2014)
\bibitem{Ayon-Beato:2000mjt}
E.~Ayon-Beato and A.~Garcia,
Phys. Lett. B \textbf{493}, 149-152 (2000)
\bibitem{Brustein:2018web}
R.~Brustein and A.~J.~M.~Medved,
Phys. Rev. D \textbf{99}, 064019 (2019)



\bibitem{I. Dymnikova}
I. Dymnikova, M. Korpusik,
 Phys. Lett. B {\bf{685}}(1), 12–18 (2010)









%


\end{thebibliography}
\end{document}